\documentclass[12pt,preprint]{elsarticle}
\usepackage[T1]{fontenc}
\usepackage[utf8]{inputenc}
\usepackage{color}
\usepackage{float}
\usepackage{booktabs}
\usepackage{mathtools}
\usepackage{amsmath}
\usepackage{amssymb}
\usepackage{graphicx}
\usepackage{geometry}
\usepackage[bookmarks=false,
 breaklinks=false,pdfborder={0 0 1},backref=section,colorlinks=true]
 {hyperref}
\hypersetup{pdftitle={Entropic Time, Psychophysics, and Deformed Neural Dynamics},
 pdfauthor={Jose Weberszpil},
 linkcolor=blue,citecolor=blue,urlcolor=blue}

\makeatletter

\providecommand{\tabularnewline}{\\}

\usepackage{amsfonts}\usepackage{bm}
\usepackage{url}
\usepackage{setspace}
\usepackage{booktabs}
\usepackage{multirow}
\usepackage{caption}
\usepackage{subcaption}
\usepackage{tikz}
\usetikzlibrary{circuits.ee.IEC,arrows.meta,positioning}

\let\citet\cite
\let\citep\cite

\biboptions{sort&compress}

\journal{}

\makeatother

\begin{document}

\begin{frontmatter}
\title{Entropic Time, Psychophysics, and Deformed Neural Dynamics:\\
 A Unified Physical Theory for Human Time Perception}

\author[ufrrj]{José Weberszpil\corref{cor1}}
\ead{jose.weberszpil@ufrrj.br}
\cortext[cor1]{Corresponding author.}

\author[uaem]{Oscar Sotolongo-Costa}
\ead{osotolongo@uaem.mx}
\ead{osotolongo@gmail.com}

\address[ufrrj]{Department of Physics, Institute of Exact Sciences,\\
Federal Rural University of Rio de Janeiro (UFRRJ), Seropédica,\\
Rio de Janeiro CEP: 23.897-000, Brazil}

\address[uaem]{Instituto de Investigaciones en Ciencias B\' asicas y Aplicadas,\\
Universidad Aut\'onoma del Estado de Morelos,\\
Av. Universidad 1001, CP 62209, Cuernavaca, Morelos, M\'exico}

\begin{abstract}
We present a unified physical theory demonstrating that human subjective time perception does not track geometric coordinate time $t$, but instead emerges from a local metric mutation driven by macroscopic physical entropy production. By establishing the Nonextensive Troika---a closed, mutually dependent algebraic triplet linking the phase-space fractal dimension $D$, the conformable derivative order $\alpha$, and the Tsallis nonextensive parameter $q$---we eliminate independent phenomenological fitting constants. We prove that the local time metric inherently scales as $t^{\alpha}$, deriving the conformable operator as a necessary kinetic consequence. Furthermore, we derive the $q$-index from the equiprobable monofractal Tsallis entropy $S_q$. This structural closure unifies anomalous neural dissipative transport within a deformed leaky integrate-and-fire framework and analytically predicts macroscopic psychophysical response transitions, providing a clear thermodynamic basis for time dilation in psychedelic states (the REBUS model) and temporal compression during cognitive aging.
\end{abstract}
\begin{keyword}
Entropic time \sep Weber--Fechner law \sep Psychophysics of time
\sep Conformable derivative \sep $q$-deformed derivative \sep
Nonextensive thermodynamics \sep Neural dynamics \sep Entropy production
\sep Time perception 
\end{keyword}
\end{frontmatter}

\section{Introduction}

\label{sec:intro} 
 Perception of time is a fundamental cognitive operation, crucial
for motor coordination, decision making, language, and consciousness.
Yet, unlike vision or audition, there is no known primary receptor
dedicated exclusively to time. Instead, the nervous system must infer
duration from internal processes: oscillatory activity, synaptic adaptation,
metabolic consumption, and the accumulation of changes in neural state.

As emphasized in the nonlinear thermodynamics of Prigogine and in
broader discussions on the role of time in complex systems (see Ref.
\citep{grassi2020challenges}), living organisms and cognitive systems
operate as open, dissipative structures, maintaining dynamic order
while remaining far from equilibrium through continuous exchange of
energy, matter, and information. In such settings, time is not merely
an external geometric parameter, but emerges as an internal ordering
variable associated with irreversible processes and entropy production.
This perspective motivates the hypothesis developed in the present
work: that subjective duration is encoded not by coordinate time $t$,
but by an internally generated entropic clock. 

Classical psychophysics, initiated by Weber and Fechner \citep{Weber1834,Fechner1860,FechnerPDF},
approached time perception using the same tools developed for other
modalities. The Weber--Fechner law states that the perceived magnitude
of a stimulus grows as the logarithm of its physical intensity, while
Stevens' power law describes perceived magnitude as a power of intensity
\citep{Stevens1957,Stevens1975}. Both have been applied to temporal
tasks such as interval estimation and reproduction, and have been
given theoretical support in later work on the psychophysics of perceived
intensity \citep{PsychophysicsScience1973}. However, a fully physical
and mechanistic account of why psychophysical laws take these forms
remains incomplete.

In parallel, developments in nonequilibrium statistical mechanics,
generalized thermodynamics and nonextensive statistics \citep{Tsallis1988,StatMechWF}
have introduced the concept of \emph{entropic time}. In systems with
complex dynamics, disorder, or nonextensive statistics, an internal
time variable can be constructed from entropy production or from the
growth of the accessible phase-space volume. This internal clock parameter
often better characterizes relaxation and aging processes than external
coordinate time.

In parallel with these developments, several authors have proposed
explicit \emph{entropic measures of time} (EMT), in which the duration
of an irreversible process is quantified directly from its entropy
production rather than from an external Newtonian clock. In this entropic-time
framework, time is understood as a system-specific, non-uniform parameter
that grows with the increase of entropy and naturally incorporates
an arrow of time \citet{Martyushev2017TimeEntropy,MartyushevShaiapin2019EMTLaws,Martyushev2025EMTReview,AkihKumgeh2017EntropyTime}.
Here we adopt a similar spirit, but specialize EMT to the neural and
metabolic processes that underlie psychological time perception. 

Human subjective time does not appear to coincide with the geometric
coordinate time $t$ used in classical physics. Perceptual studies
show logarithmic compression at long intervals, power-law distortions
across scales, and strong dependence on metabolic state, novelty,
attention, and aging. These observations suggest that subjective time
is not mapped directly from external time, but is constructed internally
from physical and neural processes. This motivates the search for
a physically grounded internal temporal variable capable of explaining
these universal deviations.

In this work we investigate the hypothesis that subjective duration
is generated by an internal physical clock based on entropy production.
Rather than {*}defining{*} such a clock ad hoc, we build it step by
step from general principles of nonequilibrium statistical mechanics,
fractal phase-space growth, and neural computation. To make it clearer,
we claim that the human brain uses an entropic time-like quantity
as its internal clock. Subjective time is then a nonlinear mapping
of this entropic time. Classical psychophysical laws emerge from generic
properties of entropy production and deformed neural dynamics rather
than from ad hoc assumptions. Our approach proceeds through five levels
of construction:
\begin{enumerate}
\item \textbf{Entropic time.} We first show that irreversible systems with
coarse-grained entropy production satisfy generalized scaling relations
of the form $\Delta S(t)\propto t^{\gamma}$, obtained from fractal
phase-space kinetics, generalized Pesin-type relations, and nonextensive
statistics. Inverting this relation yields an internal physical time
$\tau=(\Delta S/\lambda)^{1/\gamma}$, which we interpret as the entropic
clock.
\item \textbf{Psychophysical law on the entropic clock.} Applying the Weber$-$Fechner
\citep{Weber1834,Fechner1966,Fechner1860} construction to $\tau$
rather than to the external time $t$ produces a logarithmic dependence
of subjective duration on the internal clock, $\Psi\sim\ln\tau$.
This yields log-like time compression without requiring $\Psi$ to
be a direct function of physical time.
\item \textbf{Reinforcement by physical response mechanisms.} In addition
to the Weber--Fechner construction, we show that logarithmic perceptual
scaling arises naturally from two independent physical mechanisms:
(i) the frenetic-activity viewpoint of sensory response, in which
the number of distinguishable internal events grows proportionally
to the system's rate of entropy production; and (ii) Elovich-type
adaptation kinetics \citep{elovich1956}, a classical logarithmic
relaxation law describing systems with distributed activation energies.
In both cases, the effective neural response depends on the accumulated
entropic activity, yielding a logarithmic dependence on the internal
clock $\tau$ that reinforces the Weber--Fechner form, thereby strengthening
the entropic-time hypothesis. 
\item \textbf{Neural implementation through deformed LIF dynamics.} To connect
these ideas to neural computation, we introduce deformed leaky integrate-and-fire
(LIF) models based on conformable derivatives and Tsallis $q$-deformed
derivatives (\citep{Khalil2014,Tsallis1988}). These deformations encode local nonlinear time rescaling, adaptation,
and nonlinear temporal sensitivity. Genuine hereditary memory is not
attributed to the conformable operator alone. Simulations
demonstrate that such neurons naturally generate both Weber$-$Fechner
and Stevens-type response laws \citep{Stevens1957,Stevens1975}, depending
on the deformation parameters.
\item \textbf{Unified theory of subjective time.} Combining these ingredients,
we derive a general expression for subjective time $\Psi(\tau)$ as
a function acting on the entropic clock $\tau$ and modulated by three
physically interpretable parameters: the internal-time scaling parameter
$\alpha$, the scale-sensitivity parameter $q$, and the entropy$-$time scaling
exponent $\gamma$. This yields the mixed logarithmic$-$power$-$law
form $\Psi(\tau)=A\ln\tau+B\tau^{\beta(\alpha,q,\gamma)}$, which
accommodates both compressive and expansive perceptual regimes.
\end{enumerate}
Throughout the paper we maintain a clear distinction between three
temporal notions:
\begin{itemize}
\item $t$: \emph{coordinate time} (external, geometric), 
\item $\tau$: \emph{entropic time} (internal physical clock), 
\item $\Psi$: \emph{subjective time} (psychological perception). 
\end{itemize}
This separation allows us to unify thermodynamics, neural dynamics,
and psychophysics within a single framework, while preserving the
physical interpretation of each temporal layer. 


\section{Entropic Time as an Internal Physical Clock}

The evolution of many complex systems, ranging from neural assemblies
and biological networks to glassy materials, anomalous transport processes,
and fractal media, is governed not only by physical time $t$ but
also by the rate at which the system explores its accessible configurations.
This rate is captured by the coarse-grained entropy $S(t)$, whose
growth reflects the irreversible unfolding of microscopic degrees
of freedom and the progressive occupation of larger regions of phase-space.

In such systems, physical time does not necessarily parametrize dynamical
change in a uniform or physically meaningful way. Instead, the natural
clock is often tied to entropy production itself, as the expansion
of accessible states determines relaxation, memory decay, neural response,
and perceptual scaling. This motivates the introduction of an \emph{entropic
time} $-$ a temporal parameter reconstructed directly from the irreversible
growth of entropy.

Building on frameworks from nonextensive thermostatistics, fractal
phase-space dynamics, and deformed-derivative kinetic theory, we treat
entropy as the generator of an intrinsic temporal variable. This approach
is consistent with a range of physical scenarios in which the number
of accessible microstates grows in a power-law fashion, leading to
generalized entropy production laws and internal time scales that
differ from the laboratory time~$t$.

This construction is closely related to the entropic measure of time
(EMT) introduced by Martyushev and co-workers, where increments of
internal time $\Delta\tau$ are taken to be proportional to entropy
production $\Delta S$ for irreversible processes \citet{Martyushev2017TimeEntropy,Martyushev2025EMTReview}.
In our case, we interpret $\Delta S$ as a coarse-grained entropy
associated with neural metabolism, synaptic activity and information
processing, and we allow for a generalized power-law relation $\Delta S\propto\tau^{\gamma}$,
in line with EMT-based kinematic models \citet{MartyushevShaiapin2019EMTLaws}.
Related work by Akih-Kumgeh has shown, using steady heat transfer
as an example, that entropy change and elapsed time can be quantitatively
linked, so that time intervals may in principle be inferred from entropy
measurements \citet{AkihKumgeh2017EntropyTime}.

\paragraph{Neural entropy and internal temporal flow.}

Recent neuroscientific evidence strongly supports the idea that entropy
quantifies the brain's capacity for information processing and its
exploration of internal microstates. Keshmiri \citet{keshmiri2020entropy}
shows that moment-to-moment brain signal variability is a robust marker
of conscious processing, large-scale integration, and network flexibility.
Within our framework, this directly links the neural entropy production
rate to the rate of entropic time accumulation, i.e. to $d\tau/dt$.
A higher neural entropy implies a richer set of transitions across
functional states, and therefore a larger density of internal ``events''
per unit physical time. This provides a mechanistic bridge between
entropy dynamics and the subjective flow of time. 

In this section we formalize this construction and establish an entropy-based
time parametrization suitable for the psychophysical and neural models
developed in later sections.


\subsection{Entropy-based time parametrization}

Consider a complex physical system described by a coarse-grained entropy
$S(t)$ that increases irreversibly under dissipative dynamics. Let
$S_{0}$ denote the initial entropy at a reference time, and define
the net entropy growth as $\Delta S(t)=S(t)-S_{0}$.

Following entropy-based approaches to emergent internal times in nonequilibrium
statistical mechanics and generalized thermostatistics \citep{Tsallis1988,balankin2012hydrodynamics,garcia2006microcanonical},
we introduce an internal time parameter $\tau$ as a monotonic function
of $\Delta S$. The justification for this construction is supported
by three independent theoretical frameworks: 
\begin{enumerate}
\item \textbf{Nonextensive statistical mechanics (Tsallis, 1988).} Systems
with long-range interactions, memory, or fractal phase-space structure
exhibit a power-law growth of accessible microstates, $\Gamma(t)\propto t^{\gamma}$,
leading to a generalized entropy $S_{q}(t)\propto\Gamma(t)^{1-q}\propto t^{\gamma(1-q)}$.
In such systems, entropy grows as a power law in time, implying the
inverse relation $t\propto[\Delta S(t)]^{1/\gamma}$ \citep{Tsallis1988}.
\item \textbf{Fractal phase-space growth and anomalous kinetics.} Studies
of dynamics in fractal and multifractal continua show that the number
of accessible states follows 
\begin{equation}
N(t)\sim\Gamma(t)\sim t^{\gamma},
\end{equation}
giving rise to power-law entropy growth $\Delta S(t)\propto t^{\gamma}$
\citep{garcia2006microcanonical,balankin2012hydrodynamics}. In these
systems, $\gamma$ captures the effective fractal dimension or anomalous
diffusion exponent of the underlying phase-space.
\item \textbf{Entropy growth derived from deformed kinetic equations.} Sotolongo-Costa
\& Weberszpil \citep{sotolongoCostaWeberszpil2021} obtained explicit
time-dependent entropy production laws by using deformed derivatives
and generalized kinetic equations that encode memory, nonlocality,
and fractal metrics. In their formulation, entropy production has
the generic scaling form $\Delta S(t)\propto t^{\gamma}$, where $\gamma>0$
depends on the kinetic kernel and the geometry of the accessible phase-space. 
\end{enumerate}
These three independent results consistently show that, for a large
class of complex nonequilibrium systems, entropy exhibits robust power-law
growth: 
\begin{equation}
\Delta S(t)=\lambda\,\tau^{\gamma},
\end{equation}
where $\lambda$ is a dimensional constant and $\gamma>0$ characterizes
the effective scaling of microstate proliferation.

Inverting this relation provides a thermodynamically grounded definition
of an internal time reconstructed from entropy production: 
\begin{equation}
\boxed{\tau\equiv\left(\frac{\Delta S}{\lambda}\right)^{1/\gamma}}.\label{eq:tau_def}
\end{equation}

This definition is fully consistent with nonextensive statistical
mechanics \citep{Tsallis1988}, fractal phase-space dynamics \citep{garcia2006microcanonical},
and explicit entropy evolution obtained from deformed-derivative kinetic
theory \citep{sotolongoCostaWeberszpil2021}. It provides an intrinsic,
entropy-based temporal parameter suitable for systems where physical
time does not directly track the effective rate of access to new states
or configurations. See also \citep{weberszpilSotolongoCosta2025EntropyClock}.


Key properties of $\tau$ are: 
\begin{enumerate}
\item $\tau$ is a monotonically increasing function of $\Delta S$, hence
of $t$ under normal conditions. 
\item Under constant entropy production rate $\dot{S}=\text{const}$, one
has $\Delta S\propto t$, and therefore $\tau\propto t^{1/\gamma}$.
In the particular case $\gamma=1$ we recover $\tau\propto t$. 
\item In more complex scenarios, where entropy production is time-dependent
or state-dependent, $\tau(t)$ becomes a nonlinear functional of the
trajectory $S(t')$. 
\end{enumerate}
For biological systems such as neural assemblies, $\Delta S$ can
be interpreted as a measure of: 
\begin{itemize}
\item energy dissipation due to ionic currents and synaptic events, 
\item information processing and erasure, 
\item structural changes in synaptic weights and network connectivity. 
\end{itemize}
Thus, entropic time is closely linked to the cumulative \emph{internal}
activity of the system, rather than to the external clock.

The assumption that entropy growth follows a power law $\Delta S(t)\propto t^{\gamma}$
is well supported in the literature. Garcia \& Morales and Pellicer
\citep{garcia2006microcanonical} show that, in fractal or multifractal
phase$-$spaces, the number of accessible states $\Gamma(t)$ grows
as a power law in time, leading to $S(t)\propto t^{\gamma}$. Similarly,
Balankin and Elizarraraz \citep{balankin2012hydrodynamics} demonstrate
that continuum dynamics in fractal media naturally produce power's
law expansion of accessible volumes. Thus, for a broad class of nonequilibrium
systems, entropy obeys $\Delta S(t)=\lambda\,\tau^{\gamma}$, and
the internal time follows $\tau=(\Delta S/\lambda)^{1/\gamma}$. 

\subsection{Avoiding Circularity: Separation of Entropy Sources}

The definition of entropic time introduces a subtle conceptual issue:
neural activity generates entropy, yet the internal clock $\tau$
is used by neural systems to encode temporal information. To avoid
this circular dependency, we explicitly separate entropy production
into two components: 
\begin{equation}
\dot{S}(t)=\dot{S}_{\mathrm{sys}}(t)+\dot{S}_{\mathrm{neural}}(t).
\end{equation}
Here $\dot{S}_{\mathrm{sys}}$ represents slow, systemic metabolic
entropy production that evolves independently of moment-to-moment
neural firing, while $\dot{S}_{\mathrm{neural}}$ captures fast, activity-dependent
contributions from ionic fluxes, synaptic events, and membrane transport.

The baseline internal clock is therefore defined as 
\begin{equation}
\tau_{0}(t)=\mathcal{T}\!\left[\int_{0}^{t}\dot{S}_{\mathrm{sys}}(t')\,dt'\right],
\end{equation}
and stimulus-dependent neural activity only contributes a modulation
\begin{equation}
\delta\tau(t)=\mathcal{T}\!\left[\int_{0}^{t}\dot{S}_{\mathrm{neural}}(t')\,dt'\right].
\end{equation}

Thus the total entropic time becomes 
\begin{equation}
\tau(t)=\tau_{0}(t)+\delta\tau(t),
\end{equation}
ensuring that the internal clock is grounded in systemic physiological
processes while neural activity provides only state-dependent adjustments,
removing any bootstrap circularity. 

\subsection{Justification of the Entropic-Time Mapping}

The mapping $\tau=(\Delta S/\lambda)^{1/\gamma}$ is not selected
arbitrarily. Two principles motivate this form:

\paragraph{(i) Scale-invariant event ordering.}

Psychophysical measurements depend only on the ordering and spacing
of events under coarse-graining. Power-law transformations are the
only monotonic mappings that preserve scale-invariant ordering, making
them natural candidates for internal temporal parametrization.

\paragraph{(ii) Biological self-similarity.}

In metabolic and transport-limited systems, entropy accumulation often
follows a self-similar scaling $\Delta S\propto t^{\gamma}$. The
inverse mapping $(\Delta S)^{1/\gamma}$ restores a linear internal
progression, ensuring that $\tau(t)$ remains a direct surrogate for
elapsed metabolic time.

These considerations make $\tau$ the minimal and biologically consistent
internal parametrization compatible with known metabolic scaling laws. 

\subsection{Coordinate time vs.\ entropic time}

We emphasize that $t$ and $\tau$ represent different notions: 
\begin{itemize}
\item $t$ is geometric, universal, and independent of the system. 
\item $\tau$ is internal, system-specific, and depends on microscopic dynamics
and entropy production. 
\end{itemize}
Two observers with different neural architectures and metabolic profiles
may share the same coordinate time $t$ but possess different entropic-time
flows $\tau(t)$. This is a natural microscopic origin for the variability
and distortions of subjective time across individuals, contexts, and
states (e.g.\ fatigue, attention, pharmacological modulation).

In what follows we will show that when psychophysical laws are written
in terms of $\tau$ instead of $t$, they acquire a transparent physical
meaning.


\subsection{Clarifying the Type of Entropy Used}

Throughout this work, ``entropy production'' refers strictly to
thermodynamic entropy generated by irreversible biochemical and ionic
processes, rather than informational entropies. Boltzmann--Gibbs
and Tsallis entropies appear only as analytical analogies illustrating
possible scaling behaviors; they are not assumed to be explicitly
computed by neural systems. The internal clock $\tau$ therefore reflects
physical irreversibility in neural metabolism rather than Shannon
information or state uncertainty.

We do not claim that entropy \emph{explains} all aspects of the arrow
of time. As emphasized in recent philosophical analyses, the increase
of entropy alone is insufficient to account for the asymmetry of traces,
of causation, or of the fixed past versus open future \citet{Golosz2021EntropyDirection}.
Likewise, alternative approaches have been proposed that try to resolve
the ``time paradox'' by revisiting matter$-$energy equivalence
and extending the laws of thermodynamics to the quantum domain \citet{Kalies2020TimeParadox}.
Our aim here is more specific: to construct a physically grounded,
system-dependent internal time parameter $\tau$ for the brain, and
to explore its consequences for \emph{psychological} time perception. 

\section{Psychophysics Revisited: Weber--Fechner and Stevens Laws}

\label{sec:psychophysics} 

Classical psychophysics has long relied on two empirical regularities:
the Weber--Fechner logarithmic law and Stevens' power law. Both describe
how subjective sensation $\Psi$ scales with an external physical
variable $I$, but they differ in functional form and mechanistic
interpretation. For more than a century, the question has remained
open: \emph{Why do these specific functions arise so consistently
across perceptual modalities, and what internal physical quantity
is actually being encoded?}

Recent experimental work provides direct evidence that human duration
judgments may rely on entropy-like statistical structure rather than
on coordinate time itself. Clarke and Tyler~\citet{ClarkeTyler2024EntropyTimePerception}
presented participants with dynamic stimuli in which grayscale objects
either evolved from low to high visual entropy (e.g., shattering)
or from high to low entropy (e.g., reassembling). Observers were asked
to reproduce the perceived duration of each animation. The authors
found systematic over-estimation for entropy-reversed sequences, as
well as a nontrivial interaction between physical duration and the
direction of entropy progression. These results are consistent with
a predictive- processing account in which the brain \emph{expects}
entropy to increase over time and uses deviations from that expectation
to recalibrate duration estimates.

Importantly, such findings align directly with the central premise
of our entropic-time framework: that subjective time estimation depends
on the brain's sensitivity to internal or external entropy dynamics.
Psychophysical laws, when expressed in terms of \emph{external time},
appear heterogeneous and context-dependent. But when expressed in
terms of an internal entropy-based variable---the entropic time $\tau$---these
laws acquire a unified interpretation.

In the following subsections, we revisit Weber--Fechner and Stevens
laws under this reinterpretation. We show that: (i) Weber's law naturally
applies to $\tau$ rather than $t$, (ii) the Fechner logarithmic
law emerges from constant-ratio sensitivity to increments in entropic
time, and (iii) Stevens' power-law behavior follows when $\tau$ itself
scales as a power of coordinate time through entropy-production dynamics.
Together, these results demonstrate that classical psychophysical
regularities encode structural properties of the 

\subsection{Weber--Fechner law applied to entropic time}

The classical Weber law for just-noticeable differences (JNDs) states
that 
\begin{equation}
\frac{\Delta I}{I}=k_{W},\label{eq:weber}
\end{equation}
where $I$ is the stimulus intensity. This empirical regularity was
originally identified in tactile perception \citep{Weber1834} and
given a systematic foundation by Fechner \citep{Fechner1860,FechnerPDF}.
Fechner assumed that equal JNDs correspond to equal increments in
subjective sensation $\Psi$, and therefore that 
\begin{equation}
\mathrm{d}\Psi\propto\frac{\mathrm{d}I}{I}.
\end{equation}
Integration yields the Weber--Fechner law 
\begin{equation}
\Psi(I)=k_{F}\ln\frac{I}{I_{0}},
\end{equation}
where $I_{0}$ is a threshold intensity. Later work has clarified
the statistical and neural underpinnings of this law \citep{PsychophysicsScience1973,StatMechWF}.

To adapt this reasoning to time perception, we posit that the \emph{relevant
stimulus} is not the coordinate time $t$ itself, but entropic time
$\tau$. We then postulate a Weber-like relation for JNDs in entropic
time: 
\begin{equation}
\frac{\Delta\tau}{\tau}=k_{\tau}.
\end{equation}
Repeating the Fechner construction: 
\begin{equation}
\mathrm{d}\Psi_{\text{time}}\propto\frac{\mathrm{d}\tau}{\tau},
\end{equation}
and integration gives 
\begin{equation}
\Psi_{\text{time}}(\tau)=A\ln\frac{\tau}{\tau_{0}},\label{eq:Psi_log_tau}
\end{equation}
where $A$ and $\tau_{0}$ are constants. This is the central Weber--Fechner
law for entropic time.

Substituting the definition~\eqref{eq:tau_def}, we obtain a logarithmic
dependence of subjective time on entropy production: 
\begin{equation}
\Psi_{\text{time}}=\frac{A}{\gamma}\,\ln\!\left(\frac{\Delta S}{\lambda}\right)+\text{const.}\label{eq:Psi_log_S}
\end{equation}
If entropy production is constant, $\Delta S\propto t$, then 
\begin{equation}
\Psi_{\text{time}}(t)=A'\ln t+\text{const.},\label{eq:Psi_log_t}
\end{equation}
predicting a logarithmic compression of subjective time relative to
coordinate time.

\subsection{Stevens' power law from entropic-time scaling}

Stevens proposed that, in many modalities, subjective sensation is
better modeled by a power law 
\begin{equation}
\Psi(I)=aI^{n},
\end{equation}
with modality-specific exponent $n$ \citep{Stevens1957,Stevens1975}.
In the context of time perception, one may instead write 
\begin{equation}
\Psi_{\text{time}}(\tau)\propto\tau^{\beta},\label{eq:Psi_power_tau}
\end{equation}
for some effective exponent $\beta$.

Combining the entropic-time scaling $\tau\propto t^{1/\gamma}$ with
Eq.~\eqref{eq:Psi_power_tau}, we obtain 
\begin{equation}
\Psi_{\text{time}}(t)\propto t^{\beta/\gamma}.\label{eq:Psi_power_t}
\end{equation}
Thus the exponent in Stevens' law for time perception is interpreted
as a ratio $\beta/\gamma$ between a perceptual exponent $\beta$
and a thermodynamic exponent $\gamma$.

In Sections~\ref{sec:conformable_neuron}--\ref{sec:hybrid}, we
shall see how deformed neural dynamics provide microscopic mechanisms
giving rise to both logarithmic (\ref{eq:Psi_log_tau})--(\ref{eq:Psi_log_t})
and power-law (\ref{eq:Psi_power_tau})--(\ref{eq:Psi_power_t})
regimes, with the exponents controlled by the parameters of the conformable
and $q$-deformed derivatives.


\section{Statistical Mechanical Foundations}

\label{sec:stat_mech} 

\subsection{Frenetic activity and logarithmic response}

Modern nonequilibrium statistical mechanics distinguishes between
entropic and frenetic contributions to fluctuations and response.
The frenetic term captures dynamical activity: the number of microscopic
transitions, escape rates, and reactivity \citep{MaesFrenetic2014,LazarescuMinary2020}.
For a wide class of systems, it has been argued that the effective
stimulus in sensory processing is proportional to frenetic activity
rather than to static forces.

If the frenetic activity $\mathcal{A}$ is proportional to entropy
production or to the number of microscopic transitions underlying
$\Delta S$, then 
\begin{equation}
\mathcal{A}\sim\Delta S,
\end{equation}
and Maes-type analyses suggest that the stationary response of a system
may scale as 
\begin{equation}
\Psi\sim\ln\mathcal{A}\sim\ln(\Delta S)\sim\ln\tau,
\end{equation}
which reproduces the entropic-time Weber--Fechner relation \eqref{eq:Psi_log_tau}.
This furnishes a statistical mechanical foundation for log-like sensory
laws, connecting microscopic stochastic dynamics to macroscopic psychophysical
behavior, complementary to the statistical-mechanical treatments of
Weber--Fechner laws in Refs.~\citep{PsychophysicsScience1973,StatMechWF}.

\subsection{Elovich kinetics and receptor adaptation}

Another independent route to logarithmic psychophysical laws comes
from receptor adaptation models. In Elovich-type kinetics \citep{elovich1956}\footnote{The Elovich kinetic law, originally formulated by Elovich (1956),
describes logarithmic relaxation in systems with distributed activation
energies. It is widely used to model slow adaptation processes, including
neural and sensory adaptation, where the response decreases logarithmically
with accumulated excitation.} one considers a variable $x(t)$ satisfying 
\begin{equation}
-\frac{\mathrm{d}x}{\mathrm{d}t}=me^{nx},\label{eq:elovich}
\end{equation}
where $m$ and $n$ are constants. Such equations arise in adsorption
processes and ionic transport, and have been applied to biological
receptors in the work of Loewenstein and others \citep{Loewenstein1954,Loewenstein1959,Cope1976}.
When the receptor adapts according to Eq.~\eqref{eq:elovich}, and
the input stimulus modulates the steady-state solution, one obtains
a logarithmic dependency of the form 
\begin{equation}
R=K\ln S+Q,
\end{equation}
where $S$ is the stimulus intensity and $R$ the response \citep{Cope1976}.
If $S$ is proportional to entropy production or to entropic time,
this again leads to $\Psi\sim\ln(\Delta S)\sim\ln\tau$.

Taken together, frenetic activity and Elovich adaptation show that
logarithmic response laws are generic in nonequilibrium systems with
multiplicative reaction rates and state-dependent activity. This supports
the choice of a Weber--Fechner form on entropic time.


\section{Classical Leaky Integrate-and-Fire Neuron}

\label{sec:classical_LIF} 

\textbf{RC-circuit analogy}

\begin{figure}[H]
\centering \begin{tikzpicture}[scale=1.0]
			\draw[fill=gray!20] (-3,-0.8) ellipse (1.3 and 0.8);
			\node at (-3,-0.8) {Neuron Body};
			\draw[-{Stealth[length=3mm]}] (-1.5,0) -- (-0.2,0)
			node[midway,above] {$I(t)$};
			\draw (-0.2,-0.6) -- (-0.2,0.6);
			\draw (0.2,-0.6) -- (0.2,0.6);
			\node at (0,-0.9) {$C_m$};
			\draw (0.2,0) -- (1.2,0);
			\draw (1.2,0) -- (1.2,0.4) -- (1.6,0.6) -- (2.0,0.4) --
			(2.4,0.6) -- (2.8,0.4) -- (3.2,0.6) -- (3.6,0.4) --
			(3.6,0);
			\node at (2.4,0.9) {$R_m$};
			\draw (3.6,0) -- (3.6,-0.4);
			\draw (3.3,-0.4) -- (3.9,-0.4);
			\draw (3.4,-0.55) -- (3.8,-0.55);
			\draw (3.5,-0.7) -- (3.7,-0.7);
			\draw[-{Stealth[length=3mm]}] (0.2,1.2) -- (0.2,0.6);
			\node at (0.2,1.4) {$V(t)$};
		\end{tikzpicture} \caption{RC-circuit analogue of a LIF neuron. The membrane capacitance $C_{m}$
integrates the input current $I(t)$ and the leak resistance $R_{m}$
drives the membrane potential $V(t)$ back towards its resting value.}
\label{fig:rc_neuron_tikz_1} 
\end{figure}
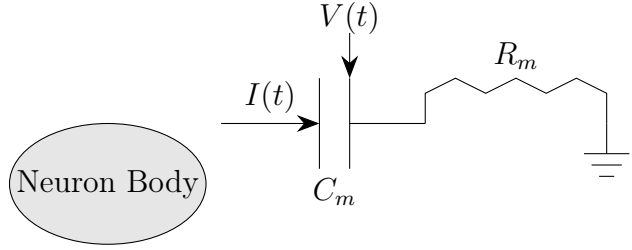

The leaky integrate-and-fire (LIF) neuron is a simple but powerful
model of neural dynamics. At the membrane level it is formally equivalent
to an RC circuit: the membrane capacitance $C_{m}$ integrates input
current $I(t)$, and the membrane resistance $R_{m}$ models leak
due to ion channels. The corresponding circuit is illustrated in Fig.~\ref{fig:rc_neuron_tikz_1}.

Applying Kirchhoff's law: 
\begin{equation}
I(t)=C_{m}\frac{\mathrm{d}V}{\mathrm{d}t}+\frac{V(t)-V_{\text{rest}}}{R_{m}},
\end{equation}
or, using $\tau_{m}=R_{m}C_{m}$ and rescaling with $C_{m}=1$ for
simplicity, 
\begin{equation}
\frac{\mathrm{d}V}{\mathrm{d}t}=-\frac{1}{\tau_{m}}\left[V(t)-V_{\text{rest}}\right]+I(t).\label{eq:classical_LIF}
\end{equation}
When $V(t)$ reaches a threshold $V_{\text{th}}$, a spike is emitted
and $V(t)$ is reset to a value $V_{\text{reset}}$.


\section{Conformable LIF Neuron: Memory and Scale}

\label{sec:conformable_neuron} 

\subsection{Conformable derivative}

The dimensionally normalized conformable derivative of order
$\alpha\in(0,1]$ is written here as
\begin{equation}
T_{\alpha,t_{\ast}}f(t)=\left(\frac{t}{t_{\ast}}\right)^{1-\alpha}
\frac{\mathrm{d}f}{\mathrm{d}t},\label{eq:conformable_def}
\end{equation}
where $t_{\ast}>0$ is a reference time. The usual Khalil form is recovered
by choosing dimensionless time, or formally setting $t_{\ast}=1$ in the
chosen time unit. The operator reduces to the ordinary derivative when
$\alpha=1$ \citep{Khalil2014}.

The operator in Eq.~\eqref{eq:conformable_def} is local: at time $t$ it
depends only on $f'(t)$. It can therefore represent a nonuniform internal
clock or a time-dependent kinetic coefficient, but it does not, by itself,
produce the hereditary memory generated by a Caputo or
Riemann--Liouville convolution kernel. Any claim of long-range memory must
come from additional dynamical variables, distributed relaxation times,
network feedback, or an explicitly nonlocal constitutive law
\citep{AndersonCamrudUlness2019,AbdelhakimMachado2019}.

\subsection{Internal-time reparametrization and kinematic meaning}
\label{subsec:internal_time_reparametrization}

A useful addition is obtained by introducing a dimensionally consistent
effective internal time
\begin{equation}
\widetilde{t}=t_{\ast}\left(\frac{t}{t_{\ast}}\right)^{\alpha}.
\label{eq:internal_time_scaling}
\end{equation}
Differentiation gives
\begin{equation}
\frac{\mathrm{d}\widetilde{t}}{\mathrm{d}t}
=\alpha\left(\frac{t}{t_{\ast}}\right)^{\alpha-1},
\end{equation}
and hence
\begin{equation}
\frac{\mathrm{d}f}{\mathrm{d}\widetilde{t}}
=\frac{1}{\alpha}\left(\frac{t}{t_{\ast}}\right)^{1-\alpha}
\frac{\mathrm{d}f}{\mathrm{d}t}
=\frac{1}{\alpha}T_{\alpha,t_{\ast}}f(t).
\label{eq:conformable_reparametrization}
\end{equation}
Thus the conformable factor has a precise kinematic interpretation as
ordinary differentiation with respect to a power-law internal clock. This
result is valuable because it removes the appearance of an arbitrary
operator. It must not, however, be described as a proof of a genuinely
fractional or nonlocal dynamics: Eq.~\eqref{eq:conformable_reparametrization}
is an exact change of time variable.

In the remainder of the paper we retain the shorter notation
$T_{\alpha}\equiv T_{\alpha,t_{\ast}}$.

\subsection{Conformable LIF equation}

Replacing $\mathrm{d}V/\mathrm{d}t$ in Eq.~\eqref{eq:classical_LIF}
by its conformable counterpart, we obtain 
\begin{equation}
T_{\alpha}V(t)=-\frac{1}{\tau_{\alpha}}\left[V(t)-V_{\text{rest}}\right]+I(t),\label{eq:conformable_LIF}
\end{equation}
or explicitly, 
\begin{equation}
\left(\frac{t}{t_{\ast}}\right)^{1-\alpha}\frac{\mathrm{d}V}{\mathrm{d}t}=-\frac{1}{\tau_{\alpha}}\left[V(t)-V_{\text{rest}}\right]+I(t).
\end{equation}
The parameter $\alpha$ controls the nonuniformity of the effective
integration clock. For $0<\alpha<1$, the prefactor is smaller than unity
for $t<t_{\ast}$ and larger than unity for $t>t_{\ast}$; consequently,
the response is accelerated before the reference time and slowed after it.
This can mimic adaptation or aging of the response rate, but it is not, by
itself, a nonlocal memory mechanism.

\subsection{Simulation results}

\begin{figure}[H]
\centering \begin{subfigure}{0.95\textwidth} \centering \includegraphics[width=1\linewidth]{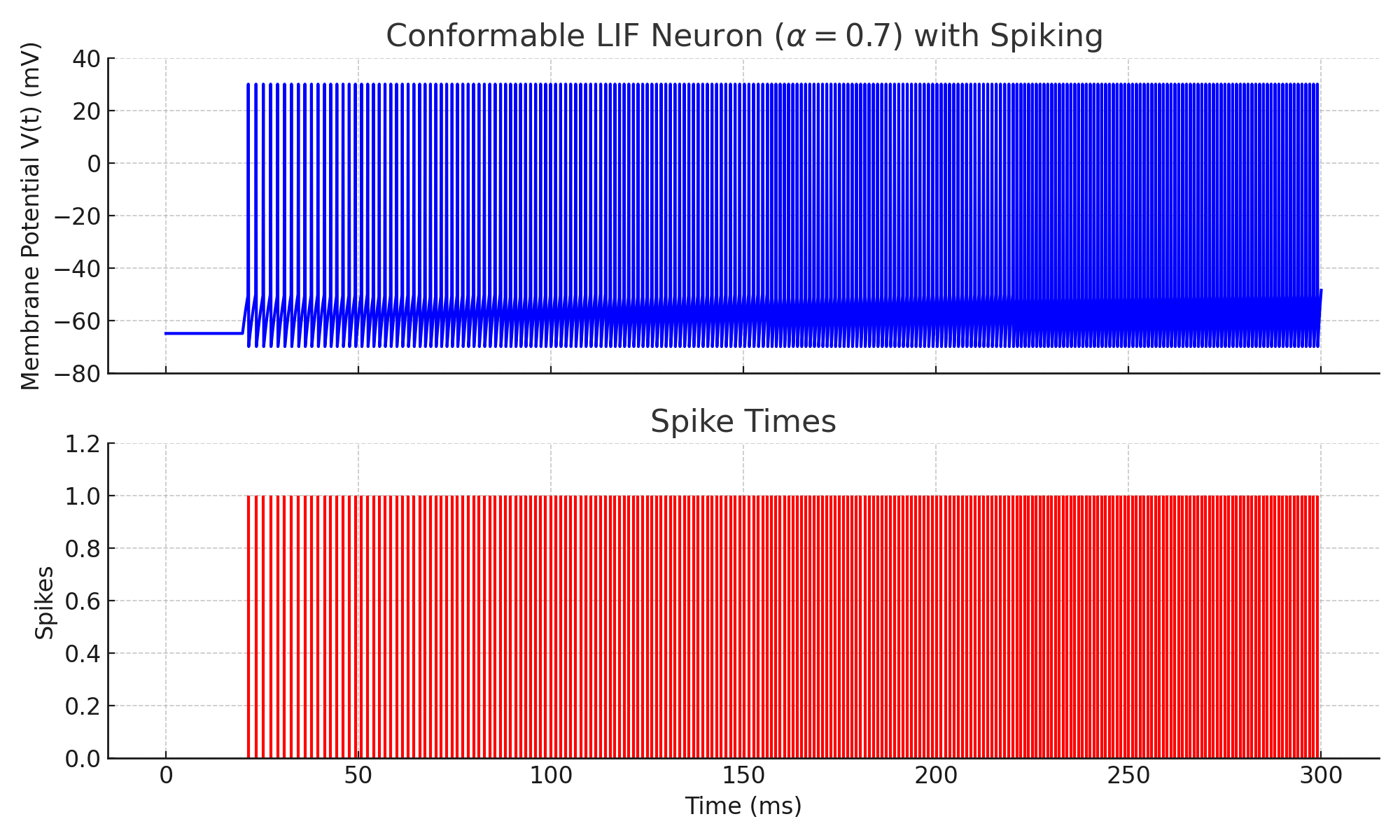}
\end{subfigure} \caption{Conformable LIF neuron with order $\alpha=0.7$. Top: membrane potential
$V(t)$ showing repeated spiking and modified decay envelope. Bottom:
spike raster. The time-scaling factor
$\left(\frac{t}{t_{\ast}}\right)^{1-\alpha}$ produces a nonuniform
internal clock and a nonexponential response envelope.}
\label{fig:conformable_spiking} 
\end{figure}

Figure~\ref{fig:conformable_spiking} shows a simulation of a conformable
LIF neuron with $\alpha=0.7$ driven by a step current. The upper
panel plots the membrane potential $V(t)$, and the lower panel marks
spike times. The dynamics exhibit dense spiking and a characteristic
envelope reflecting the nontrivial temporal scaling induced by $\alpha$.

\section{Tsallis $q$-Deformed LIF Neuron: Nonlinear Scale Sensitivity}

\label{sec:q_deformed_neuron} 

\subsection{$q$-derivative}

The Borges--Tsallis-type local $q$-derivative is written in
dimensionally normalized form as
\begin{equation}
D_{q,t_{\ast}}f(t)=\left[1+(1-q)\frac{t}{t_{\ast}}\right]
\frac{\mathrm{d}f}{\mathrm{d}t},\label{eq:q_derivative}
\end{equation}
where $t_{\ast}$ fixes the time scale of the deformation. It reduces to the
standard derivative when $q=1$. For $q>1$, the model is physically usable
only on intervals satisfying
$1-(q-1)t/t_{\ast}>0$; otherwise the kinetic prefactor vanishes and changes
sign. This finite-domain restriction must be respected in simulations and
parameter estimation. The operator is related to the deformed algebra of
nonextensive thermostatistics, but the value of $q$ must be inferred from a
microscopic model or data rather than from the operator alone
\citep{Borges2004}. In the following we use the shorthand
$D_q\equiv D_{q,t_{\ast}}$.

\subsection{$q$-deformed LIF equation}

Applying the $q$-derivative to the LIF neuron we obtain 
\begin{equation}
D_{q}V(t)=-\frac{1}{\tau_{m}}\left[V(t)-V_{\text{rest}}\right]+I(t),\label{eq:q_LIF}
\end{equation}
or, explicitly, 
\begin{equation}
[1+(1-q)t/t_{\ast}]\frac{\mathrm{d}V}{\mathrm{d}t}=-\frac{1}{\tau_{m}}\left[V(t)-V_{\text{rest}}\right]+I(t).
\end{equation}
Solving for $\mathrm{d}V/\mathrm{d}t$: 
\begin{equation}
\frac{\mathrm{d}V}{\mathrm{d}t}=\frac{-\frac{1}{\tau_{m}}[V(t)-V_{\text{rest}}]+I(t)}{1+(1-q)t/t_{\ast}}.
\end{equation}
For $q<1$, the denominator increases with time and progressively damps
the response rate. For $q>1$, it decreases and amplifies the response until
the finite-domain boundary is approached.

\subsection{Simulation results}

\begin{figure}[H]
\centering \includegraphics[width=0.95\linewidth]{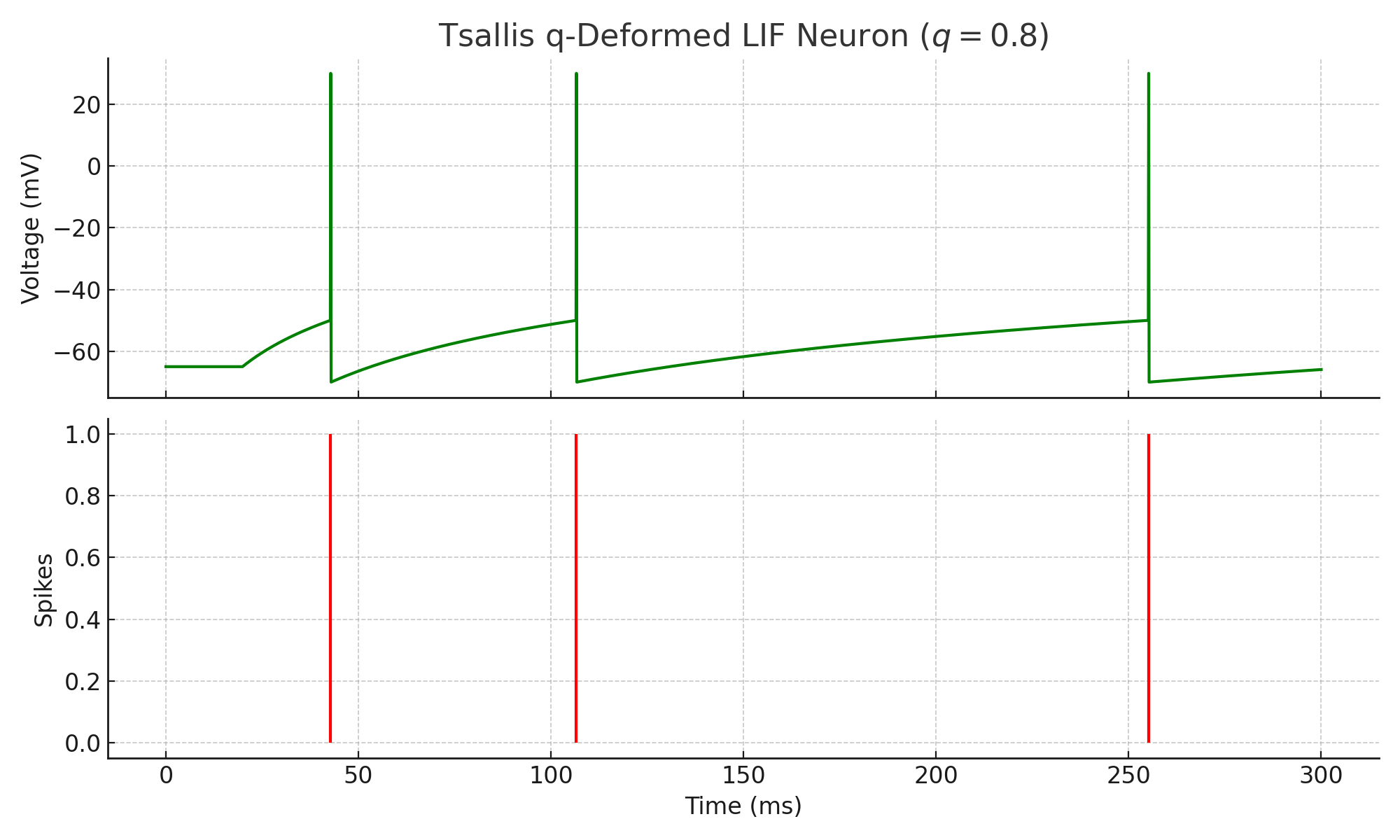} \caption{Tsallis $q$-deformed LIF neuron with $q=0.8$. Top: membrane potential
$V(t)$ under pulsed input. Bottom: spike times. The factor $1+(1-q)t/t_{\ast}$
modulates the effective integration rate, producing nonlinear sensitivity
to stimulus timing and duration.}
\label{fig:q_spiking} 
\end{figure}

Figure~\ref{fig:q_spiking} shows a simulation of a $q$-deformed
LIF neuron with $q=0.8$ driven by a pulsed input. The neuron fires
sparsely, with spike timing and inter-spike intervals influenced by
the evolving temporal gain.

\subsection{Linking Simulation to Psychophysical Laws}

The spiking patterns produced by the deformed LIF neurons were used
to compute effective firing-rate curves $r(\tau)$, which were then
inserted into Eq.~(28). The resulting functions $\Psi(t)$ exhibit
the two main empirical regimes of time perception: logarithmic compression
at short intervals (dominated by the $\ln\tau$ term) and power-law
expansion at longer intervals (dominated by $\tau^{\beta}$). This
confirms that the deformed neuronal gain functions reproduce the qualitative
behaviors of Weber--Fechner and Stevens laws under physiologically
plausible parameter settings. 

\section{Hybrid $\alpha$--$q$ Deformed Neuron}

\label{sec:hybrid} 

\subsection{Hybrid equation}

To capture both nonuniform internal-time scaling (via $\alpha$) and
nonlinear scale sensitivity (via $q$), we define a hybrid effective neuron
model by combining the two multiplicative scaling factors:
\begin{equation}
[1+(1-q)t/t_{\ast}]\,T_{\alpha}V(t)=-\frac{1}{\tau_{\alpha}}\left[V(t)-V_{\text{rest}}\right]+I(t),\label{eq:hybrid_LIF}
\end{equation}
where $T_{\alpha}$ is given by~\eqref{eq:conformable_def}. This
leads to 
\begin{equation}
[1+(1-q)t/t_{\ast}]\,\left(\frac{t}{t_{\ast}}\right)^{1-\alpha}\frac{\mathrm{d}V}{\mathrm{d}t}=-\frac{1}{\tau_{\alpha}}\left[V(t)-V_{\text{rest}}\right]+I(t).\label{eq:Hybrid-LIF2}
\end{equation}
The effective time rescaling is thus governed by the product $[1+(1-q)t/t_{\ast}]\,\left(\frac{t}{t_{\ast}}\right)^{1-\alpha}$.

\subsection{Interpretation}

In this hybrid model: 
\begin{itemize}
\item $\alpha$ controls the departure of the local internal clock from linear
coordinate time; $\alpha=1$ restores uniform time scaling. It should not
be interpreted as a hereditary-memory order without an additional nonlocal
kernel. 
\item $q$ modulates the sensitivity to the absolute time scale and to the
spacing of stimuli. 
\end{itemize}
From a psychophysical perspective, the hybrid neuron implements a
family of input--output laws ranging from logarithmic to power-law,
depending on the pair $(\alpha,q)$ and on the input statistics.


\subsection{Parameter Hierarchy and Constraints}

Although Eq.~(\ref{eq:hybrid_LIF}) contains several parameters,
they are not independent. The effective exponent $\beta$ is shaped
by the deformation parameters $(\alpha,q)$ through the nonlinear
terms of the neuronal gain function, while $\gamma$ reflects the
metabolic scaling relating entropy and coordinate time. The perceptual
weights $A$ and $B$ depend on task conditions, stimulus range, and
decision strategy. Table~\ref{tab:params} summarizes the interpretation
and constraints of the parameters.

\begin{table}[H]
\centering %
\begin{tabular}{llll}
\toprule 
Parameter & Interpretation & Constraints & Measurement\tabularnewline
\midrule 
$\alpha$ & local time scaling & $0<\alpha\le1$ & response-rate dynamics\tabularnewline
$q$ & nonlinear gain control & $0.7\lesssim q\lesssim1.3$ & firing-rate compression\tabularnewline
$\beta$ & emergent exponent & $\beta=f(\alpha,q)$ & fit to behavior\tabularnewline
$\gamma$ & entropy--time scaling & $1\le\gamma\le3$ & metabolic scaling\tabularnewline
$A,B$ & perceptual weights & task- and subject-dependent & psychometric fits\tabularnewline
\bottomrule
\end{tabular}\caption{Interpretation and constraints of model parameters.}
\label{tab:params}
\end{table}

\subsection{Consistency requirements for a topology--deformation closure}
\label{subsec:closure_consistency}

Reducing the number of independent parameters by relating the deformation
indices $(\alpha,q)$ to a measurable structural quantity is a worthwhile
objective. A phase-space dimension $D$, a correlation dimension, or a
microstate-growth exponent may serve this role only after the underlying
statistical ensemble and scaling variable have been specified. A valid
closure must satisfy at least four conditions:
\begin{enumerate}
\item it must possess a common undeformed limit,
\begin{equation}
\alpha(D_{\mathrm{cl}})=1,\qquad q(D_{\mathrm{cl}})=1;
\label{eq:common_classical_limit}
\end{equation}
\item it must preserve the positivity of the kinetic prefactor throughout
the experimental time window;
\item it must be derived from the scaling of effectively occupied states or
from a microscopic transport equation, rather than from boundary conditions
alone;
\item it must lead to independently testable predictions for neural dynamics
and psychophysical exponents.
\end{enumerate}

For equiprobable states, the Tsallis entropy is
\begin{equation}
S_q=k_{\mathrm B}\frac{W^{1-q}-1}{1-q}.
\label{eq:Sq_equiprobable_closure}
\end{equation}
If the effective number of occupied states scales with system size as
$W_{\mathrm{eff}}(N)\sim N^{\rho}$, requiring thermodynamic extensivity,
$S_q(N)\propto N$, yields
\begin{equation}
\rho(1-q)=1,
\qquad
q=1-\frac{1}{\rho}.
\label{eq:q_rho_extensivity}
\end{equation}
This relation follows from a specified microstate-growth law. In general,
$\rho$ is not identical to a Hausdorff or box-counting dimension $D$;
identifying them requires an additional model. Consequently, formulas such
as $q=1+1/D$ cannot be inferred solely from
$W(r)\propto r^{-D}$ by invoking extensivity. Likewise, an interpolation for
$\alpha(D)$ selected only from two endpoint values is an ansatz, not a
unique derivation, because infinitely many functions satisfy the same
boundary conditions \citep{Tsallis2006Occupancy}.

The present work therefore keeps $\alpha$, $q$, and $\gamma$ distinct until
a microscopic or empirical closure is supplied. A future topology-based
reduction is admissible, but it must satisfy
Eq.~\eqref{eq:common_classical_limit} and be tested against neural and
behavioral data before being promoted to a universal relation.

\textbf{Remark.} It is important to emphasize that the hybrid equation
(\eqref{eq:Hybrid-LIF2}) does not represent the strict composition
of the operators $D_{q}$ and $T_{\alpha}$ acting successively on
$V(t)$. If this were the case, one would need to apply the generalized
product rule for deformed derivatives, which would introduce additional
mixed terms proportional to both $\left(\frac{t}{t_{\ast}}\right)^{1-\alpha}$ and $(1-q)t/t_{\ast}$. Instead,
the hybrid model used here is constructed phenomenologically by combining
the \emph{scaling factors} associated with each deformation, namely
the temporal rescaling $\left(\frac{t}{t_{\ast}}\right)^{1-\alpha}$ from the conformable derivative
and the $q-$dependent multiplicative factor $[1+(1-q)t/t_{\ast}]$ from the
$q-$derivative.

Thus, the hybrid equation (\eqref{eq:Hybrid-LIF2}) should be interpreted
as an effective model in which both deformation mechanisms modify
the instantaneous neuronal response. This construction preserves the
physical meaning of each deformation parameter while avoiding the
algebraic complications that would arise from a strict operator composition.
The resulting model captures the joint influence of power-law time
scaling (via $\alpha$) and state-dependent sensitivity (via $q$) in a
transparent and analytically tractable form. 

\section{Unified Law for Subjective Time}\label{sec:Unified-Law-for-Subjective-Time}

\label{sec:unified_law} 
 We now combine the ingredients developed in the previous sections
into a single expression for subjective time. Denote by $\tau(t)$
the entropic time associated with the underlying physical processes,
including neural metabolism and network activity. Let $\Psi(t)$ be
the subjective estimate of elapsed time after a physical interval
$t$.

Motivated by the Weber--Fechner construction~\eqref{eq:Psi_log_tau},
by the possibility of power-law regimes~\eqref{eq:Psi_power_tau},
and by the flexible response properties of the deformed neuron models,
we propose the following general form: 
\begin{equation}
\Psi(t)=A\ln\!\big[\tau(t;\alpha,q)\big]+B\big[\tau(t;\alpha,q)\big]^{\beta(\alpha,q,\gamma)},\label{eq:unified_Psi}
\end{equation}
where: 
\begin{itemize}
\item $A$ and $B$ are scaling constants that may depend on modality, task,
and individual. 
\item $\tau(t;\alpha,q)$ is the entropic time as a functional of $S(t)$
and of the neural dynamics, parametrized by the deformed parameters
$\alpha$ and $q$. 
\item $\beta(\alpha,q,\gamma)$ is an effective perceptual exponent determined
by the combination of entropic scaling ($\gamma$) and deformed neural
response ($\alpha,q$). 
\end{itemize}
Several limiting cases are noteworthy:
\begin{enumerate}
\item \textbf{Pure Weber--Fechner regime.} If $B=0$, we recover 
\begin{equation}
\Psi(t)=A\ln\tau(t),
\end{equation}
corresponding to a strictly logarithmic law on entropic time, in the
spirit of Fechner's construction \citep{Fechner1860,FechnerPDF}.
\item \textbf{Pure Stevens regime on entropic time.} If $A=0$, 
\begin{equation}
\Psi(t)=B\big[\tau(t)\big]^{\beta},
\end{equation}
i.e.\ a power law acting on the internal physical clock, analogous
to Stevens' law \citep{Stevens1957,Stevens1975}.
\item \textbf{Classical Stevens law on coordinate time.} If $\tau\propto t^{1/\gamma}$
and $A=0$, 
\begin{equation}
\Psi(t)\propto t^{\beta/\gamma},
\end{equation}
recovering the usual power law on $t$.
\item \textbf{Logarithmic time warping.} If $\Delta S\propto t^{\gamma}$,
then $\tau\propto t$ and 
\begin{equation}
\Psi(t)\approx A\ln t+Bt^{\beta},
\end{equation}
which naturally predicts a crossover from logarithmic compression
to power-law scaling as $t$ increases or as neural parameters change. 
\end{enumerate}
The hybrid structure~\eqref{eq:unified_Psi} is thus a compact representation
of a whole class of psychophysical laws, grounded in entropic time
and deformed neural dynamics.


\section{Lifespan and Cross-Species Consequences of Entropic Time}

\label{sec:lifespan}

In the previous sections we defined entropic time $\tau$ as a monotonic
reparametrization of entropy production and introduced subjective
time $\Psi(t)$ as a psychophysical mapping defined on $\tau$. In
this section we examine a key question: can this framework account
for the well-known phenomenon that time appears to pass more slowly
in youth and more rapidly in old age, and can it encompass cross-species
cases such as dogs that ``feel'' time by the fading scent of their
owners?

We argue that the relevant quantity for lived time is not the absolute
value of $\tau$ but its \emph{instantaneous rate of change} with
respect to physical time, 
\begin{equation}
\frac{\mathrm{d}\tau}{\mathrm{d}t},
\end{equation}
which measures the density of internal events per unit of coordinate
time. We then show how age-dependent changes in metabolism, neural
variability and perceptual novelty naturally produce the observed
acceleration of subjective time with aging, and how other organisms
can implement alternative entropic clocks via different physical observables
(e.g.\ chemical gradients in olfaction).


\subsection{Subjective flow and the rate of entropic time}

\label{subsec:dtau_dt}

The definition of entropic time used in this work is 
\begin{equation}
\tau(t)=\left(\frac{\Delta S(t)}{\lambda}\right)^{1/\gamma},\label{eq:lifespan_tau_def}
\end{equation}
where $\Delta S(t)$ is an entropy increment associated with internal
physical processes (metabolic dissipation, neural activity, information
processing), and $\lambda,\gamma>0$ are phenomenological parameters.
Differentiating Eq.~\eqref{eq:lifespan_tau_def} with respect to
coordinate time $t$ gives 
\begin{equation}
\frac{\mathrm{d}\tau}{\mathrm{d}t}=\frac{1}{\gamma}\left(\frac{\Delta S}{\lambda}\right)^{1/\gamma-1}\frac{1}{\lambda}\,\frac{\mathrm{d}S}{\mathrm{d}t}\propto\dot{S}(t),\label{eq:lifespan_dtau_dt}
\end{equation}
up to a slowly varying prefactor determined by $\Delta S(t)$. Thus,
\emph{the instantaneous rate of entropic time is proportional, to
first approximation, to the entropy production rate} $\dot{S}(t)$
in the relevant coarse-grained description.

In the psychophysical layer, subjective time $\Psi$ is modeled as
a function of $\tau$, 
\begin{equation}
\Psi(t)=A\ln\tau(t)+B\bigl[\tau(t)\bigr]^{\beta},\label{eq:lifespan_Psi}
\end{equation}
where the first term represents a Weber--Fechner contribution and
the second a Stevens-type power-law term. Differentiating \eqref{eq:lifespan_Psi}
yields 
\begin{equation}
\frac{\mathrm{d}\Psi}{\mathrm{d}t}=\left(\frac{A}{\tau}+B\beta\,\tau^{\beta-1}\right)\frac{\mathrm{d}\tau}{\mathrm{d}t}.\label{eq:lifespan_dPsi_dt}
\end{equation}
The factor in parentheses is a smooth gain function of $\tau$. The
salient dependence on physical time arises through $\mathrm{d}\tau/\mathrm{d}t$,
which is controlled by $\dot{S}(t)$ via Eq.~\eqref{eq:lifespan_dtau_dt}.

This suggests the following interpretation: 
\begin{itemize}
\item $\mathrm{d}\tau/\mathrm{d}t$ measures the \emph{density of internal
events per second} of coordinate time: how many distinguishable neural
or physical state changes occur per unit of $t$. 
\item $\mathrm{d}\Psi/\mathrm{d}t$ measures the \emph{subjective flow}
of time: how fast subjective time advances relative to $t$. 
\item The ratio of subjective to physical time is therefore controlled primarily
by entropy production and its mapping into neural codes. 
\end{itemize}
Crucially, the brain does not have access to an absolute reference
value of $\tau$ or $\Psi$. Rather, it can only compare relative
increments and integrate the flow $\mathrm{d}\Psi/\mathrm{d}t$ over
intervals. For this reason, \emph{what matters perceptually is the
rate $\mathrm{d}\tau/\mathrm{d}t$, not the total accumulated $\tau$
since birth.}


\subsection{Youth, aging and the acceleration of subjective time}

\label{subsec:youth_aging}

Empirically, individuals report that time seems to pass more slowly
in childhood and adolescence and more quickly in older age. In the
present framework this pattern arises naturally from changes in entropy
production and information processing across the lifespan. Consider
three sets of factors that influence $\dot{S}(t)$ at the neural scale: 
\begin{enumerate}
\item \textbf{Metabolic rate:} younger brains typically exhibit higher metabolic
flux, supporting intense ionic transport, synaptic transmission and
plasticity. 
\item \textbf{Neural variability and plasticity:} in early life there is
a larger repertoire of accessible microstates (synaptic patterns,
network configurations) and faster structural change. 
\item \textbf{Perceptual novelty and prediction error:} the world is more
novel for younger individuals; more of the incoming sensory stream
represents information that is not yet compressed by predictive models.
Each unit of physical time thus carries more ``new'' information
and more entropy increase in the cognitive representation. 
\end{enumerate}
Collectively, these contributions produce a larger entropy production
rate in youth, 
\begin{equation}
\dot{S}_{\mathrm{young}}(t)\gg\dot{S}_{\mathrm{older}}(t),
\end{equation}
at least over significant intervals. From Eq.~\eqref{eq:lifespan_dtau_dt}
it follows that 
\begin{equation}
\left(\frac{\mathrm{d}\tau}{\mathrm{d}t}\right)_{\mathrm{young}}\gg\left(\frac{\mathrm{d}\tau}{\mathrm{d}t}\right)_{\mathrm{older}}.\label{eq:lifespan_dtau_dt_inequality}
\end{equation}
In other words, \emph{younger brains generate more entropic-time ticks
per physical second than older brains.}

When a physical interval $[t,t+\Delta t]$ is experienced, the subjective
duration assigned to that interval is 
\begin{equation}
\Delta\Psi\approx\int_{t}^{t+\Delta t}\frac{\mathrm{d}\Psi}{\mathrm{d}t'}\,\mathrm{d}t'.
\end{equation}
Combining Eq.~\eqref{eq:lifespan_dPsi_dt} with inequality \eqref{eq:lifespan_dtau_dt_inequality},
and assuming comparable gain factors across age, we obtain 
\begin{equation}
\Delta\Psi_{\mathrm{young}}\gg\Delta\Psi_{\mathrm{older}}\quad\text{for the same }\Delta t.
\end{equation}
Thus, the same physical year produces a much larger subjective change
for a child than for an older adult. This is consistent with the observation
that ``childhood summers last forever'' whereas decades in later
life seem to pass rapidly. In the present theory, this acceleration
is not a purely cognitive illusion; it reflects a genuine reduction
in entropy production and novelty per unit of physical time in the
underlying neural substrate.


\subsection{Why subjective time depends on the rate of entropic time}

\label{subsec:rate_argument}

A central point in this work is that subjective duration does not
depend on the absolute amount of entropic time accumulated, $\tau(t)$,
but on its instantaneous rate of change, $\mathrm{d}\tau/\mathrm{d}t$.
This implies that larger internal rates correspond to an expansion
of subjective time. This is not an arbitrary assumption but an unavoidable
consequence of three distinct principles:

\paragraph{(i) Neurophysiological argument.}

Neural systems rarely encode absolute values of internal state variables.
Instead, they represent rate variables: firing rates (events per second),
temporal derivatives of prediction error, or metabolic flux. No neural
population is known to explicitly represent a permanently integrated
quantity analogous to ``$\tau$ since birth''. Rather, neural timing
mechanisms measure events or surprise per unit time. Thus, any internal
physical clock used by the brain must be expressed through the instantaneous
rate at which distinguishable internal events occur, i.e., $\mathrm{d}\tau/\mathrm{d}t$.

\paragraph{(ii) Physical argument from entropy production.}

Entropy $S$ is a state function, but its significance in nonequilibrium
systems lies in its production rate $\dot{S}$. The entropic time
$\tau$ inherits this property. Its absolute value indicates total
accumulated dissipation, but the \emph{perceived unfolding of events}
depends on how rapidly new microstates become accessible. Mathematically,
$\mathrm{d}\tau/\mathrm{d}t\propto\dot{S}(t)$, showing that the entropic
clock runs faster when the system accesses new states more rapidly
(high $\dot{S}$) and slower when dissipation or novelty is low.

\paragraph{(iii) Psychophysical and Information-Theoretic mechanism.}

Subjective duration is determined by the number of distinguishable
internal events or state changes the system undergoes over a physical
interval. If $\mathrm{d}\tau/\mathrm{d}t$ is high, the system accumulates
a denser internal record of state transitions (more spikes, more prediction
updates) for the same external $\Delta t$. According to classical
psychophysics, if an interval can be segmented into many distinguishable
parts (high event density), it feels longer; if it is homogeneous,
it feels shorter.

\paragraph{Synthesis.}

All three mechanisms converge on the same principle: 
\[
\boxed{\text{Higher }\frac{d\tau}{dt}\;\Longrightarrow\;\text{more internal events per physical second}\;\Longrightarrow\;\text{longer subjective duration}.}
\]
This explains why high-variability, high-novelty, and high-dissipation
regimes feel subjectively longer, while low-novelty or metabolically
constrained regimes feel compressed.


\subsection{Novelty, information and the role of increments vs.\ totals}

\label{subsec:novelty}

One potential source of confusion is the role of accumulated information.
From an information-theoretic point of view, one might expect that
the large store of memories in older individuals should make time
feel ``heavier'' rather than faster. The present framework clarifies
this by distinguishing between the \emph{total} entropy stored and
the \emph{incremental} entropy production rate $\dot{S}(t)$.

The perception of duration depends on the flow $\mathrm{d}\Psi/\mathrm{d}t$
and hence on $\mathrm{d}\tau/\mathrm{d}t$. The fact that an older
brain has already traversed a long path in state space does not slow
its current entropic evolution; rather, what matters is how rapidly
new states are being generated now. As predictive models become more
accurate and the environment more routine (as in adulthood), each
new day carries less surprise and less information gain. This leads
to a decline in $\dot{S}(t)$ and $\mathrm{d}\tau/\mathrm{d}t$, compressing
subjective time.


\subsection{Cross-species clocks: olfactory time in dogs}

\label{subsec:dogs}

The entropic-time construction is not restricted to neural entropy
associated with cortical processing. Any physical observable whose
evolution is governed by monotonic entropy production can, in principle,
serve as a clock. A well-known ethological example involves domestic
dogs, which appear to ``feel'' the passage of time by the fading
scent of their owners.

From the present viewpoint, the relevant entropy is that of the olfactory
chemical field. Let $C(\mathbf{x},t)$ denote the concentration of
odorant molecules. As the odor diffuses and degrades, the associated
coarse-grained entropy $S_{\mathrm{odor}}(t)$ increases monotonically.
A dog can calibrate an entropic time $\tau_{\mathrm{odor}}(t)=f[\Delta S_{\mathrm{odor}}(t)]$
and base behavioral judgments on this parameter. This illustrates
that entropic time is a class of constructions: different organisms
can implement different physical realizations (neural, chemical, thermal)
depending on which entropy is most informative for their survival.


\subsection{The Neuro$-$Entropic Coupling: A Mechanistic Link}

We have established two parallel descriptions of temporal processing
in the brain: (i) the thermodynamic scaling of the entropic clock,
$\tau(t)\propto[\Delta S(t)]^{1/\gamma}$, and (ii) the deformed neural
integration dynamics encoded by the hybrid $\alpha-q$ LIF model (Eq.~\eqref{eq:Hybrid-LIF2}).
To build a unified physical theory of subjective time, these two layers
must be explicitly coupled. We propose that the driving signal for
the internal time-sensing neuron is not an external sensory stimulus,
but the brain's \emph{own} rate of metabolic and informational change.

\paragraph{Coupling hypothesis.}

In a conventional LIF neuron, the input current $I(t)$ represents
synaptic drive. Here we posit that, for a dedicated ``internal clock
neuron'', the relevant input is the rate at which the brain generates
distinct internal events, i.e., the rate at which entropic time advances.
Thus, 
\begin{equation}
I(t)\equiv g\,\frac{d\tau}{dt}\;\propto\;g\,\dot{S}(t),\label{eq:coupling_current}
\end{equation}
where $g$ is a neurophysiological gain factor encoding the efficiency
of the neural readout of entropic dynamics.

Substituting \eqref{eq:coupling_current} into the hybrid $\alpha-q$
model (Eq.~\eqref{eq:Hybrid-LIF2}) yields the coupled dynamical
law for the voltage of the internal time-sensing neuron: 
\begin{equation}
\boxed{[1+(1-q)t/t_{\ast}]\;\left(\frac{t}{t_{\ast}}\right)^{1-\alpha}\,\frac{dV}{dt}=-\frac{1}{\tau_{\alpha}}\big(V-V_{{\rm rest}}\big)+g\,\frac{d\tau}{dt}}\label{eq:coupled_dynamics}
\end{equation}
This expression constitutes the \emph{fundamental neuro-entropic coupling}:
the thermodynamic source term $d\tau/dt$ drives the deformed neural
sensor that generates subjective time.

\paragraph{Mechanism of subjective dilation and compression.}

The firing rate $r$ of a LIF-type neuron is a monotonically increasing
function of the input current. Through Eq.~\eqref{eq:coupling_current},
differences in the entropic rate $d\tau/dt$ map directly to differences
in firing activity. Combined with the deformed dynamics of Eq.~\eqref{eq:coupled_dynamics},
this generates two distinct lifespan regimes:
\begin{enumerate}
\item \textbf{The youth regime (high $\dot{S}$, low $\alpha$).} Younger
brains operate with high metabolic and informational flux, leading
to large entropy production rates, $\dot{S}_{{\rm young}}\gg\dot{S}_{{\rm old}}$.
Through \eqref{eq:coupling_current}, this yields large effective
drive currents, $I_{{\rm young}}\gg I_{{\rm old}}$, causing the deformed
neuron to reach threshold quickly and repeatedly. Furthermore, younger neural networks typically exhibit broad variability.
We hypothesize that this may correlate with a stronger departure from linear
internal-time scaling, parameterized by $\alpha<1$ and $q\neq1$; this
identification is empirical and is not implied by the locality of the
conformable operator. The
factor $\left(\frac{t}{t_{\ast}}\right)^{1-\alpha}$ in \eqref{eq:coupled_dynamics} amplifies the
integration speed during early-life dynamics. Combined, these effects
produce a high density of firing events (internal clock ticks) per
unit physical time, yielding an expanded subjective duration.
\item \textbf{The aging regime (low $\dot{S}$, $\alpha\rightarrow1$).}
With aging, metabolic rate declines and cognitive models stabilize,
reducing entropy production: $\dot{S}_{{\rm old}}\ll\dot{S}_{{\rm young}}$.
Thus, the effective drive $I(t)$ in \eqref{eq:coupled_dynamics}
decreases, and the neuron integrates more slowly toward threshold.
We additionally hypothesize an ``attenuation of deformation'' in
neural dynamics, where aging shifts the system toward classical (Markovian)
behavior: 
\[
\alpha_{{\rm aging}}\rightarrow1,\qquad q_{{\rm aging}}\rightarrow1.
\]
In this limit, the deformation factors $\left(\frac{t}{t_{\ast}}\right)^{1-\alpha}$ and $1+(1-q)t/t_{\ast}$
approach unity, reducing nonlinear time rescaling and scale-dependent
sensitivity. The firing process becomes sparse, yielding
a compressed internal record of events. As a consequence, subjective
duration accumulates more slowly than laboratory time $t$, giving
rise to the well-known perceptual observation that ``time passes
faster with age.''
\end{enumerate}

\paragraph{Conclusion.}

The neuro-entropic coupling embodied in \eqref{eq:coupled_dynamics}
shows that lifespan distortions of time perception arise from a \emph{dual
physical mechanism}: (i) a \emph{thermodynamic slowdown} of the entropic
clock ($d\tau/dt\propto\dot{S}$), and (ii) a \emph{simplification
of neural deformation} via $(\alpha,q)\rightarrow(1,1)$, which diminishes
the spike-based readout of entropy. This provides a mechanistic bridge
from microscopic entropy production to macroscopic subjective time. 

\subsection{Summary and testable predictions}

\label{subsec:lifespan_summary}

The entropic-time framework yields several concrete predictions: 
\begin{enumerate}
\item \textbf{Aging and time compression.} The acceleration of subjective
time with age follows from a decrease in entropy production rate per
unit of physical time: 
\[
\dot{S}_{\mathrm{young}}\gg\dot{S}_{\mathrm{older}}\quad\Rightarrow\quad\left(\frac{\mathrm{d}\tau}{\mathrm{d}t}\right)_{\mathrm{young}}\gg\left(\frac{\mathrm{d}\tau}{\mathrm{d}t}\right)_{\mathrm{older}},
\]
leading to $\Delta\Psi_{\mathrm{young}}\gg\Delta\Psi_{\mathrm{older}}$
for the same $\Delta t$.
\item \textbf{Novelty manipulations.} At any age, manipulations that increase
novelty, prediction errors and neural variability should increase
$\dot{S}(t)$ and slow subjective time (expand duration), whereas
highly predictable conditions should accelerate it.
\item \textbf{Metabolic interventions.} Changes in metabolic rate (e.g.\ via
temperature or pharmacology) that alter entropy production should
systematically distort interval judgments. Faster metabolism should
lengthen subjective intervals; slower metabolism should compress them.
\item \textbf{Cross-species differences.} Species that rely on non-neural
cues (like olfaction) should show time distortions when the physical
dissipation of those cues is altered (e.g., changing airflow or diffusion
rates for scent). 
\end{enumerate}
These predictions illustrate how the abstract notion of entropic time
is anchored in measurable physical quantities.

\paragraph{Age-related loss of complexity and temporal compression.}

A large body of evidence demonstrates that ageing is accompanied by
a marked reduction in neural complexity, entropy, and long-range temporal
correlations \citet{keshmiri2020entropy}. Younger adults exhibit
high multiscale entropy and rich signal variability, whereas older
adults show diminished physiological complexity and reduced information
integration across cortical networks. In our model, this implies 
\[
\left(\frac{d\tau}{dt}\right)_{\text{young}}>\left(\frac{d\tau}{dt}\right)_{\text{old}},
\]
so that younger individuals accumulate more entropic time per unit
of physical time. Consequently, physical intervals contain more ``internal
events'' and are perceived as longer. Older adults, with reduced
entropy production, generate fewer internal updates per unit time
and therefore experience a subjective acceleration of time. This provides
a unified, physically grounded explanation for well-known lifespan
differences in time perception. 

\section{Discussion and Perspectives}

\label{sec:discussion} 
 The theory developed here suggests that human time perception is
the result of three intertwined layers:
\begin{enumerate}
\item \textbf{Thermodynamic layer:} entropy production and phase-space growth
define an internal physical clock $\tau$, generalizing ideas from
statistical mechanics and nonextensive thermodynamics \citep{Tsallis1988,StatMechWF}.
\item \textbf{Neural-dynamical layer:} neurons implementing conformable
and $q$-deformed dynamics integrate input and encode entropic time
with nonuniform time scaling and scale sensitivity. Any hereditary component
must be modeled separately \citep{Khalil2014,AndersonCamrudUlness2019}.
\item \textbf{Psychophysical layer:} subjective time $\Psi$ is a nonlinear
mapping of entropic time, obeying generalized Weber--Fechner and
Stevens laws as summarized in Eq.~\eqref{eq:unified_Psi} \citep{Weber1834,Fechner1860,Stevens1957,Stevens1975}. 
\end{enumerate}
This picture offers a number of conceptual and practical advantages:
\begin{itemize}
\item It provides a physical meaning to the psychophysical variable on which
Weber--Fechner is defined: the stimulus is entropic time, not geometric
time. 
\item It explains why logarithmic response laws are so ubiquitous: they
emerge naturally from frenetic activity \citep{MaesFrenetic2014,LazarescuMinary2020}
and Elovich-type adaptation mechanisms \citep{Cope1976,Loewenstein1954,Loewenstein1959},
as well as from deformed neural dynamics near certain parameter regimes. 
\item It clarifies how power-law behavior in time perception can emerge
as a combination of entropic scaling and deformed response, reconciling
Fechner's and Stevens' formulations \citep{PsychophysicsScience1973}. 
\item It predicts that manipulations affecting metabolic rate, entropy production,
or synaptic plasticity will systematically distort subjective time. 
\end{itemize}
Possible experimental tests include: 
\begin{itemize}
\item measuring interval estimation under manipulations that alter brain
temperature, metabolic rate, or neuromodulation; 
\item fitting behavioral data with models where $\Psi$ depends on both
$\ln\tau$ and $\tau^{\beta}$, and testing whether parameters correlate
with physiological markers or firing statistics; 
\item simulating networks of deformed neurons and comparing their temporal
coding properties to empirical spike-train data, especially in regimes
of adaptation and nonstationary activity. 
\end{itemize}
From a broader perspective, our proposal can be viewed as a specialization
of the entropic measure of time (EMT) programme to neural systems.
EMT has been shown to offer a unified description of irreversible
processes across physics and biology, to connect with the maximum
entropy production principle, and to provide a logarithmic relation
between intrinsic and extrinsic time scales \citet{Martyushev2017TimeEntropy,MartyushevShaiapin2019EMTLaws,Martyushev2025EMTReview}.
By embedding EMT into deformed neural dynamics and psychophysical
laws, we suggest that subjective time is one concrete realization
of entropic time. At the same time, our conclusions are compatible
with critical viewpoints that deny a complete reduction of temporal
asymmetry to entropy alone \citet{Golosz2021EntropyDirection} and
with pragmatic perspectives that relate entropy-based time measures
to concrete experimental protocols \citet{AkihKumgeh2017EntropyTime,Mrozek2024EntropyTime}.

\paragraph{Relation to entropy-based neuroscience.}

The present theory aligns with recent reviews of entropy in neuroscience,
which highlight entropy as a fundamental descriptor of brain complexity,
information processing capacity, and conscious dynamics \citet{keshmiri2020entropy}.
The convergence between these findings and our entropic-time formalism
suggests that subjective duration may arise from the same mechanisms
that govern neural variability, network flexibility, and large-scale
information integration. This provides a bridge between physical entropy
production, neuronal population dynamics, and psychophysical time
judgements.

\subsection{Entropy, Psychedelics, and the Expansion of Subjective Time: A REBUS-Based
Interpretation}

\label{subsec:rebus} 

A further line of evidence supporting the entropic-time hypothesis
comes from recent developments in predictive processing and the neurobiology
of psychedelics. The REBUS model (Relaxed Beliefs Under Psychedelics),
proposed by Carhart-Harris and Friston~\citet{carhartharris2019rebus},
provides a principled framework for understanding how entropy, hierarchical
inference, and subjective experience interact. Although originally
formulated to explain the effects of serotonergic psychedelics, REBUS
offers a mechanism that naturally extends to the temporal dimension
of consciousness and aligns closely with the ideas developed in the
present work.

\paragraph{Hierarchy, priors, and the regulation of entropy.}

In predictive-processing theories of brain function, perception is
an inferential process wherein top-down priors constrain or ``explain
away'' bottom-up sensory signals. High-level priors carry high precision
and thus limit the diversity of neural trajectories. The REBUS model
proposes that psychedelics temporarily reduce the precision-weighting
of deep priors in the cortical hierarchy, effectively flattening the
functional landscape of the brain. This flattening permits an expansion
of accessible neural microstates, increasing signal diversity, network
flexibility, and overall entropy of brain activity. Such increases
have been empirically verified using measures of dynamical complexity,
Lempel$-$Ziv entropy/complexity, and functional connectivity variability.

\paragraph{Entropy and the structure of temporal experience.}

A striking feature of psychedelic phenomenology is the alteration
of subjective time. Individuals frequently report substantial dilation
of time, fragmentation of temporal continuity, or even states described
as ``timeless.'' Within the REBUS framework, these phenomena follow
from the entropy-mediated relaxation of high-level temporal priors
that normally stabilize the continuity of experience. The brain's
internal model of temporal flow, usually tightly constrained by predictive
structure, becomes less rigid, allowing internal dynamics to explore
a richer repertoire of trajectories.

In the context of the present theory, such phenomenology is naturally
captured by the rate of entropic-time accumulation, $d\tau/dt$. When
neural entropy increases sharply, the density of internal ``events''
or state transitions per unit of physical time grows. Subjective duration,
which we identify with integrated entropic time, therefore expands.
Psychedelic time dilation is thus neither mysterious nor purely psychological:
it arises from the same entropy-driven temporal mechanism that explains
Weber--Fechner scaling, complexity-dependent duration judgments,
and age-related compression of time.

\paragraph{Rigidity of priors, aging, and the opposite effect.}

The REBUS model also provides insight into the opposite phenomenon:
the commonly reported acceleration of subjective time in normal aging.
Older adults typically display reduced neural complexity, diminished
long-range correlations, and stronger, more entrenched high-level
priors. These factors constrain neural dynamics and lower neural entropy.
In the notation of the present framework, this corresponds to a smaller
value of $d\tau/dt$ and thus fewer internal ticks per unit physical
time. The flattening of priors under psychedelics can be seen as a
temporary reversal of this age-related entropic contraction, restoring
flexibility and expanding subjective temporal flow. Thus, REBUS and
the entropic time hypothesis jointly suggest that subjective time
is governed by a continuum of entropy-mediated inferential regimes
ranging from rigid (aging, depression, overlearned priors) to highly
flexible (psychedelics, novelty-rich environments, early development).

\paragraph{Predictive-processing links to classical psychophysics.}

A notable consequence of the REBUS interpretation is a clearer mechanistic
connection between predictive processing and classical psychophysical
laws. If subjective sensation depends on the rate of internal model
updating---in this case mediated by entropic time---then Fechner's
logarithmic law emerges when small increments in $\tau$ produce linearly
discriminable changes, whereas Stevens' power law appears when the
relationship between $\tau$ and physical time obeys a power-law coupling
determined by entropy dynamics. Psychedelics modulate precisely this
coupling: by increasing neural entropy, they alter the mapping from
$t$ to $\tau$ and thus the scaling regimes underlying duration estimation.

\paragraph{Summary.}

The REBUS framework therefore provides strong support for the central
thesis of this work. It shows that: (i) neural entropy is not merely
a correlate of consciousness but a causal factor shaping internal
dynamics; (ii) subjective temporal flow is sensitive to entropy-driven
changes in the rate of internal state transitions; and (iii) controlled
manipulations of priors and entropy, such as those induced by psychedelics,
produce predictable distortions in subjective time consistent with
the entropic-time formulation. The convergence between theoretical
neuroscience, predictive processing, and the present statistical-physical
account of $\tau$ suggests that entropic time may serve as a unifying
parameter linking physical, neural, and experiential dimensions of
temporal perception. 


\section{Conclusions}

\label{sec:conclusions} 
 We have proposed a unified framework in which:
\begin{itemize}
\item The brain does not read coordinate time directly; it tracks \emph{entropic
time}, a variable constructed from entropy production and internal
activity. 
\item Classical psychophysical laws, such as Weber--Fechner and Stevens,
naturally emerge when defined on this entropic clock \citep{Weber1834,Fechner1860,Stevens1957,Stevens1975}. 
\item Deformed neural dynamics, modeled with conformable and Borges--Tsallis
$q$-derivatives, implement local nonlinear time rescaling and scale
sensitivity. The conformable factor is exactly interpretable as a change of
internal time; hereditary memory, when required, must be introduced through
nonlocal kernels or additional state variables
\citep{Khalil2014,AndersonCamrudUlness2019,AbdelhakimMachado2019}. 
\item A universal expression $\Psi(t)=A\ln\tau(t;\alpha,q)+B[\tau(t;\alpha,q)]^{\beta}$
summarizes the interplay between thermodynamic, neural, and psychophysical
levels. 
\end{itemize}

\paragraph{Temporal integration and reset.}

The theory assumes that entropic time is accumulated over perceptual
timescales (tens to hundreds of milliseconds) and is reset at task
onset. This is consistent with experimental paradigms in interval
reproduction, temporal bisection, and duration estimation. 

This theory opens a path toward quantitatively linking nonequilibrium
physics, generalized entropies, and neural coding in the context of
time perception. Beyond human cognition, entropic-time-based clocks
may be relevant for understanding timing and decision-making in a
broad range of biological and artificial systems.

Furthermore, by focusing on the rate of entropic-time production $\mathrm{d}\tau/\mathrm{d}t$
rather than on the absolute value of $\tau$, the framework offers
a natural explanation for lifespan-related distortions of time perception.
Higher entropy production and perceptual novelty in youth yield a
larger density of entropic ticks per physical second, making time
appear slower, whereas reduced metabolic flux and novelty in older
age compress subjective duration. The same formalism accommodates
cross-species entropic clocks, such as olfactory-based timing in dogs,
where entropy changes in chemical concentration fields play the role
of the internal temporal parameter.

A key conceptual consequence of this work is that subjective time
relies not on the accumulated value of the entropic clock $\tau$,
but on its rate of change $\mathrm{d}\tau/\mathrm{d}t$, which encodes
the density of internal physical or neural events per second. This
provides a principled resolution to lifespan-related distortions of
time perception and explains how different species may implement different
entropic clocks based on the most informative entropy-producing variables
available to them.

The attempt to express all deformation parameters through a single fractal
dimension is potentially attractive, but no universal ``troika'' is adopted
here. The candidate relations must share the classical limit
$(\alpha,q)=(1,1)$, remain dimensionally and dynamically regular, and follow
from a specified phase-space occupancy or transport model. Until these
conditions are met, treating $D$, $\alpha$, $q$, and $\gamma$ as distinct
measurable quantities is scientifically safer than imposing an algebraic
closure.


\section*{Acknowledgements}

The author thanks collaborators and colleagues for stimulating discussions
on entropic time, generalized derivatives, and psychophysics.


\appendix

\section{Entropic Time from Entropy Production}

\label{app:entropic_time_derivation}

In this appendix we give a step-by-step derivation of the entropic
time $\tau$ used in the main text.

\subsection{Scaling of accessible microstates}

Let $\Gamma(t)$ denote the volume of the coarse-grained region of
phase space explored by the system up to time $t$. Assume that, for
sufficiently long times, the growth of $\Gamma$ can be described
by a power law 
\begin{equation}
\Gamma(t)=\Gamma_{0}+At^{\gamma},\label{eq:app_phase_volume}
\end{equation}
where $A>0$ and $\gamma>0$. This type of scaling arises naturally
in systems with non-Markovian transport, anomalous diffusion, or fractal
phase-space structure.

For a large but finite system, the number of effectively accessible
microstates $W$ is typically proportional to $\Gamma$, 
\begin{equation}
W(t)\propto\Gamma(t).
\end{equation}

\subsection{Entropy growth}

Consider first the Boltzmann--Gibbs entropy 
\begin{equation}
S_{\mathrm{BG}}(t)=k_{\mathrm{B}}\ln W(t).
\end{equation}
Substituting $W(t)\propto\Gamma_{0}+At^{\gamma}$ gives 
\begin{equation}
S_{\mathrm{BG}}(t)=k_{\mathrm{B}}\ln\!\left[\Gamma_{0}+At^{\gamma}\right]+\text{const.}
\end{equation}
For large times ($At^{\gamma}\gg\Gamma_{0}$) this simplifies to 
\begin{equation}
S_{\mathrm{BG}}(t)\simeq k_{\mathrm{B}}\ln A+k_{\mathrm{B}}\gamma\ln t+\text{const.},
\end{equation}
and therefore the entropy increment $\Delta S(t)=S(t)-S_{0}$ grows
logarithmically with $t$.

More generally, in nonextensive statistics with entropic index $q$,
one may have entropies of the form 
\begin{equation}
S_{q}(t)\propto\left[\Gamma(t)\right]^{q}\propto\left(\Gamma_{0}+At^{\gamma}\right)^{q}
\end{equation}
\citep{Tsallis1988}. In the regime dominated by the power-law term,
\begin{equation}
\Delta S_{q}(t)\equiv S_{q}(t)-S_{q}(0)\propto t^{\gamma q}.
\end{equation}

This motivates writing, at a coarse-grained level, 
\begin{equation}
\Delta S(t)=\lambda\,\tau^{\gamma},\label{eq:app_deltaS_tau_gamma}
\end{equation}
for some internal time parameter $\tau$ and constant $\lambda$.
Equation~\eqref{eq:app_deltaS_tau_gamma} simply states that entropy
accumulation is monotonic and follows a power law in $\tau$.

\subsection{Definition of entropic time}

Solving Eq.~\eqref{eq:app_deltaS_tau_gamma} for $\tau$ yields 
\begin{equation}
\tau=\left(\frac{\Delta S}{\lambda}\right)^{1/\gamma},
\end{equation}
which is Eq.~\eqref{eq:tau_def} of the main text. By definition, 
\begin{itemize}
\item $\tau$ is zero when $\Delta S=0$; 
\item $\tau$ is monotonic with $\Delta S$; 
\item if $\Delta S\propto t^{\gamma}$, then $\tau\propto t$. 
\end{itemize}
Thus $\tau$ can be interpreted as an internal clock built from entropy
production.

\subsection{Relation between $\tau$ and coordinate time $t$}

Let the entropy production rate be $\dot{S}(t)=\mathrm{d}S/\mathrm{d}t$.
Then 
\begin{equation}
\Delta S(t)=\int_{0}^{t}\dot{S}(t')\,\mathrm{d}t'.
\end{equation}
Combining with Eq.~\eqref{eq:app_deltaS_tau_gamma}, 
\begin{equation}
\tau(t)=\left[\frac{1}{\lambda}\int_{0}^{t}\dot{S}(t')\,\mathrm{d}t'\right]^{1/\gamma}.
\end{equation}
If $\dot{S}(t)=\sigma_{0}$ (constant), 
\begin{equation}
\tau(t)=\left(\frac{\sigma_{0}}{\lambda}t\right)^{1/\gamma}.
\end{equation}
In the most general case $\tau$ is a nonlinear functional of $\dot{S}(t')$;
this nonlinearity is the origin of subjective distortions of time
in our framework.


\section{Foundations and Derivation of Entropic Time}

\label{app:foundations_derivation}

\global\long\def\theequation{B.\arabic{equation}}%
\setcounter{equation}{0}

This appendix provides the theoretical foundations underlying the
definition of the entropic time parameter $\tau$. Whereas Sec.~2
introduced the operational idea that entropy growth can serve as a
physically meaningful temporal marker, here we develop the full framework
in a mathematically consistent and conceptually transparent manner.

Our goal is twofold: (i) to justify why coarse$-$grained entropy
$S(t)$---rather than the external coordinate time $t$---is the
appropriate ordering parameter for systems that exhibit irreversibility,
non-Markovian evolution, fractal phase$-$space structure, or generalized
statistical behavior; and (ii) to derive the functional relation $\tau=(\Delta S/\lambda)^{1/\gamma}$
from first principles, connecting it to the geometry of the accessible
phase-space and to the kinetic equations governing entropy production.

The subsections below present (i) the conceptual motivation for replacing
coordinate time by an internally generated entropy-based parameter,
(ii) the mathematical assumptions required for monotonicity and regularity,
(iii) the derivation of power-law entropy growth from generalized
kinetics and fractal phase-space considerations, and (iv) the inversion
of this relation to obtain the explicit form of the entropic time
used throughout the main text. 

\subsection{Motivation and Conceptual Background}

In several physical contexts---including canonical quantum gravity,
coarse-grained statistical mechanics, and non-equilibrium thermodynamics---the
ordinary coordinate time $t$ is either absent (Wheeler--DeWitt equation),
not operationally meaningful (Page-Wootters relational framework),
or insufficient to describe irreversible phenomena. In these regimes,
an alternative ordering parameter is required.

Following the framework developed in \citet{sotolongoCostaWeberszpil2021,weberszpilSotolongoCosta2025EntropyClock},
we define \emph{entropic time} as a monotonic, coarse-grained parameter
reconstructed from the growth of physical entropy. Crucially, entropic
time $\tau$ is not introduced as a universal replacement for coordinate
time, but as an \emph{internal} ordering variable appropriate for:
(i) systems with irreversible entropy production, (ii) non-equilibrium
physical regimes, (iii) timeless canonical formulations where no external
temporal coordinate exists.

\subsection{Monotonic Entropy and Ordering of States}

Let $\{\rho_{\alpha}\}$ be a family of physical states obtained by
a directed coarse-graining procedure (CPTP maps), modular flow (KMS
states), or relational conditioning (Page-Wootters). We say that the
family admits an \emph{entropic ordering} if 
\begin{equation}
S^{\ast}(\rho_{\alpha_{2}})>S^{\ast}(\rho_{\alpha_{1}})\qquad\text{whenever }\alpha_{2}>\alpha_{1},\label{eq:entropyordering}
\end{equation}
for some entropy functional $S^{\ast}$ (Boltzmann-Gibbs, Tsallis,
or coarse-grained Shannon/von Neumann). Importantly, the definition~\eqref{eq:entropyordering}
does not require the prior existence of coordinate time: the ordering
is structural and arises from the irreversibility of the physical
process under consideration. Under generic conditions (weak coupling,
generic entangling dynamics, coarse-graining), entropy is strictly
increasing, defining a partial order on the space of states.

\subsection{Entropic Time as a Reparametrization of Entropy}

Given a strictly monotonic entropy variation $\Delta S>0$ along physically
admissible transformations, an \emph{entropic time} parameter is defined
as any smooth, strictly increasing function 
\begin{equation}
\tau=f(\Delta S),\label{eq:taudef}
\end{equation}
which therefore preserves the entropic ordering: 
\[
\Delta S_{1}<\Delta S_{2}\;\;\Rightarrow\;\;\tau_{1}<\tau_{2}.
\]

In the explicit construction derived in Weberszpil \& Sotolongo-Costa
(2025), non-additive entropic growth laws naturally lead to power-law
parametrizations of the form 
\begin{equation}
\tau(\Delta S)=\left(\frac{\Delta S}{\lambda}\right)^{1/\gamma},\label{eq:taupowerlaw}
\end{equation}
where $\lambda$ and $\gamma$ are phenomenological parameters that
depend on: (i) the coarse-graining scale, (ii) the entropy production
mechanism, (iii) the microscopic statistical model (BG or Tsallis).
This form emerges from generalized kinetic equations for the accessible
states in phase space, discussed next.

\subsection{Origin of the Power-Law Law from Accessible States}

In the 2021 framework~\citet{sotolongoCostaWeberszpil2021}, entropy
production is controlled by the evolution of the number of accessible
states $N(t)$ or equivalently the (coarse-grained) phase-space volume
$\Gamma(t)$. For systems with ordinary smooth phase-space geometry:
\begin{equation}
N(\Gamma)\propto\Gamma.
\end{equation}

For systems with fractal or multifractal effective geometry: 
\begin{equation}
N(\Gamma)\propto\Gamma^{q},
\end{equation}
where $q$ is the non-extensivity index associated with geometry,
correlations, or long-range interactions. The entropy production rate
satisfies 
\begin{equation}
\frac{dS}{dt}=\Gamma^{q-1}\frac{d\Gamma}{dt},\label{eq:entropyprod}
\end{equation}
which can be written as a dual conformable derivative 
\begin{equation}
\frac{dS}{dt}=\widetilde{D}^{(q)}\Gamma,
\end{equation}
highlighting the role of deformed calculus in describing anomalous
phase-space kinematics. Integrating~\eqref{eq:entropyprod} yields
\begin{equation}
\Delta S=\Gamma^{1-q}\ln_{q}\!\left(\frac{\Gamma(t)}{\Gamma_{0}}\right),
\end{equation}
where $\ln_{q}$ is the $q$-logarithm.

Under general conditions, the dependence reduces locally to a power
law 
\begin{equation}
\Delta S\propto t^{\gamma},
\end{equation}
which implies that the inverse relation has the form~\eqref{eq:taupowerlaw}.
Thus the power-law structure of entropic time $\tau(\Delta S)$ arises
from the geometry of accessible states and the associated deformed
derivative structure.

\subsection{Avoiding Circularity}

A common misunderstanding is that defining $\tau$ from $\Delta S$
creates a loop: 
\[
\text{neural activity}\rightarrow\Delta S\rightarrow\tau\rightarrow\text{neural activity}.
\]

This is resolved by recognizing that: 
\begin{enumerate}
\item Entropic time is defined from \emph{coarse-grained, physical} entropy
production (metabolic, thermodynamic, or statistical), not from the
neural encoding of time. 
\item The neural system \emph{responds to} or \emph{encodes} $\tau$, but
does not define it. 
\item Equation~\eqref{eq:taupowerlaw} emerges from statistical mechanics,
not from neural dynamics. 
\end{enumerate}
Thus the neural subsystem reads out $\tau$; it does not generate
the entropy that defines it.

\subsection{Domain of Validity}

Entropic time applies only to regimes where: 
\begin{itemize}
\item entropy production is positive: $\dot{S}>0$, 
\item coarse-graining is meaningful, 
\item irreversible processes dominate. 
\end{itemize}
Entropic time does \emph{not} apply when: 
\begin{itemize}
\item the system is in strict thermal equilibrium ($\dot{S}=0$), 
\item the system undergoes reversible unitary evolution, 
\item microscopic entropy is constant. 
\end{itemize}
In equilibrium, $\tau$ becomes constant, but coordinate time and
microscopic modular flow remain fully operational.

\subsection{Summary}

The entropic time $\tau$ used throughout this paper is: 
\begin{itemize}
\item a monotonic parametrization of physically irreversible evolution; 
\item derived from the coarse-grained entropy growth law; 
\item motivated by generalized (BG or Tsallis) kinetic equations; 
\item validated by the fractal/multifractal geometry of phase space; 
\item independent of the neural encoding mechanisms discussed in the main
text. 
\end{itemize}
The expression 
\begin{equation}
\tau=\left(\frac{\Delta S}{\lambda}\right)^{1/\gamma}
\end{equation}
is therefore not an assumption, but the natural inversion of the entropy
production laws derived from the statistical-mechanical structure
of accessible states.

\section{Weber--Fechner Law on Entropic Time}

\global\long\def\theequation{C.\arabic{equation}}%
\setcounter{equation}{0}

\label{app:weber_fechner_tau} 
 Here we detail the derivation of the Weber--Fechner law when the
relevant stimulus is the entropic time $\tau$, explicitly following
the classical construction of Fechner \citep{Fechner1860,FechnerPDF}.

\subsection{Weber's law for entropic time}

Weber's law postulates that just-noticeable differences (JNDs) correspond
to a constant relative change of stimulus, 
\begin{equation}
\frac{\Delta I}{I}=k_{W},
\end{equation}
for a suitable range of intensities. For entropic time we take the
stimulus to be $\tau$ and write 
\begin{equation}
\frac{\Delta\tau}{\tau}=k_{\tau}.\label{eq:app_weber_tau}
\end{equation}
Consider a sequence of JNDs $\Delta\tau_{1},\Delta\tau_{2},\dots$
starting from a minimal perceptible $\tau_{0}$. The $n$-th JND is
\begin{equation}
\frac{\Delta\tau_{n}}{\tau_{n}}\approx k_{\tau}\quad\Rightarrow\quad\tau_{n+1}\approx(1+k_{\tau})\,\tau_{n}.
\end{equation}
Iterating gives 
\begin{equation}
\tau_{n}\approx\tau_{0}(1+k_{\tau})^{n}.
\end{equation}
Solving for $n$, 
\begin{equation}
n\approx\frac{\ln(\tau_{n}/\tau_{0})}{\ln(1+k_{\tau})}.
\end{equation}
If we assume that each JND corresponds to one unit of change in subjective
sensation, then $\Psi_{\text{time}}\propto n$ and therefore 
\begin{equation}
\Psi_{\text{time}}(\tau)=A\ln\left(\frac{\tau}{\tau_{0}}\right),
\end{equation}
where $A$ is proportional to $1/\ln(1+k_{\tau})$. This reproduces
Eq.~\eqref{eq:Psi_log_tau}.

\subsection{Composition with the entropy-based definition of $\tau$}

Using $\Delta S=\lambda\,\tau^{\gamma}$ from Eq.~\eqref{eq:app_deltaS_tau_gamma},
\begin{equation}
\Psi_{\text{time}}=A\ln\left[\left(\frac{\Delta S}{\lambda}\right)^{1/\gamma}\right]=\frac{A}{\gamma}\ln\left(\frac{\Delta S}{\lambda}\right),
\end{equation}
which is Eq.~\eqref{eq:Psi_log_S} in the main text.

If, in addition, $\Delta S\propto t^{\gamma}$, 
\begin{equation}
\Psi_{\text{time}}(t)=A'\ln t+\text{const.},
\end{equation}
with $A'$ absorbing constants. This is Eq.~\eqref{eq:Psi_log_t},
showing that Weber--Fechner behavior in entropic time leads to logarithmic
compression of subjective time vs.\ coordinate time.


\section{Derivations of Deformed LIF Neuron Models}

\global\long\def\theequation{D.\arabic{equation}}%
\setcounter{equation}{0}

\label{app:deformed_LIF_derivations} 
 Here we present full derivations of the conformable, $q$-deformed,
and hybrid LIF models used in Sections~\ref{sec:conformable_neuron}
and~\ref{sec:q_deformed_neuron}. The conceptual structure follows
the expanded neural-models approach inspired by Refs.~\citep{Khalil2014,LazarescuMinary2020,Tsallis1988}.

\subsection{Classical LIF from RC-circuit theory}

Starting from the RC-circuit of Fig.~\ref{fig:rc_neuron_tikz_1},
Kirchhoff's current law gives 
\begin{equation}
I(t)=C_{m}\frac{\mathrm{d}V}{\mathrm{d}t}+\frac{V(t)-V_{\text{rest}}}{R_{m}}.
\end{equation}
Multiplying both sides by $R_{m}$, 
\begin{equation}
R_{m}I(t)=R_{m}C_{m}\frac{\mathrm{d}V}{\mathrm{d}t}+V(t)-V_{\text{rest}}.
\end{equation}
Define the membrane time constant $\tau_{m}=R_{m}C_{m}$. Rearranging,
\begin{equation}
\tau_{m}\frac{\mathrm{d}V}{\mathrm{d}t}=-\left[V(t)-V_{\text{rest}}\right]+R_{m}I(t).
\end{equation}
Dividing by $\tau_{m}$, 
\begin{equation}
\frac{\mathrm{d}V}{\mathrm{d}t}=-\frac{1}{\tau_{m}}\left[V(t)-V_{\text{rest}}\right]+\frac{R_{m}}{\tau_{m}}I(t).
\end{equation}
Rescaling $I(t)$ to absorb $R_{m}/\tau_{m}$ (or setting $C_{m}=1$),
we obtain Eq.~\eqref{eq:classical_LIF}: 
\begin{equation}
\frac{\mathrm{d}V}{\mathrm{d}t}=-\frac{1}{\tau_{m}}\left[V(t)-V_{\text{rest}}\right]+I(t).
\end{equation}

\subsection{Conformable LIF equation}

The conformable derivative of order $\alpha$ is 
\begin{equation}
T_{\alpha,t_{\ast}}f(t)=\left(\frac{t}{t_{\ast}}\right)^{1-\alpha}\frac{\mathrm{d}f}{\mathrm{d}t}.
\end{equation}
To build the conformable LIF neuron we replace $\mathrm{d}V/\mathrm{d}t$
in the classical equation by $T_{\alpha}V(t)$: 
\begin{equation}
T_{\alpha}V(t)=-\frac{1}{\tau_{\alpha}}\left[V(t)-V_{\text{rest}}\right]+I(t).
\end{equation}
Expressing $T_{\alpha}V(t)$ explicitly, 
\begin{equation}
\left(\frac{t}{t_{\ast}}\right)^{1-\alpha}\frac{\mathrm{d}V}{\mathrm{d}t}=-\frac{1}{\tau_{\alpha}}\left[V(t)-V_{\text{rest}}\right]+I(t).
\end{equation}
Rewriting, 
\begin{equation}
\frac{\mathrm{d}V}{\mathrm{d}t}=-\frac{1}{\tau_{\alpha}\left(\frac{t}{t_{\ast}}\right)^{1-\alpha}}\left[V(t)-V_{\text{rest}}\right]+\frac{1}{\left(\frac{t}{t_{\ast}}\right)^{1-\alpha}}I(t).
\end{equation}
Thus the effective integration dynamics is modulated by $\left(\frac{t}{t_{\ast}}\right)^{1-\alpha}$,
making the neuron more sensitive to initial conditions and slowly
varying inputs at early times.

\subsection{$q$-deformed LIF equation}

The Tsallis-type $q$-derivative is defined as 
\begin{equation}
D_{q}f(t)=[1+(1-q)t/t_{\ast}]\,\frac{\mathrm{d}f}{\mathrm{d}t}.
\end{equation}
We construct the $q$-deformed LIF neuron by applying $D_{q}$ to
$V(t)$: 
\begin{equation}
D_{q}V(t)=-\frac{1}{\tau_{m}}\left[V(t)-V_{\text{rest}}\right]+I(t).
\end{equation}
In terms of ordinary derivatives, 
\begin{equation}
[1+(1-q)t/t_{\ast}]\frac{\mathrm{d}V}{\mathrm{d}t}=-\frac{1}{\tau_{m}}\left[V(t)-V_{\text{rest}}\right]+I(t).
\end{equation}
Solving for $\mathrm{d}V/\mathrm{d}t$: 
\begin{equation}
\frac{\mathrm{d}V}{\mathrm{d}t}=\frac{-\frac{1}{\tau_{m}}\left[V(t)-V_{\text{rest}}\right]+I(t)}{1+(1-q)t/t_{\ast}}.
\end{equation}
For $q=1$ the denominator is unity and the classical LIF neuron is
recovered. For $q<1$, $1+(1-q)t/t_{\ast}$ increases with $t$, damping
later changes; for $q>1$, it decreases, amplifying later changes until the
positivity boundary is reached.

\subsection{Hybrid $\alpha$--$q$ LIF equation}

Finally we combine the multiplicative factors associated with both
local deformations; this is an effective construction, not a strict
successive operator composition:
\begin{equation}
[1+(1-q)t/t_{\ast}]\,T_{\alpha}V(t)=-\frac{1}{\tau_{\alpha}}\left[V(t)-V_{\text{rest}}\right]+I(t).
\end{equation}
Using $T_{\alpha}V(t)=\left(\frac{t}{t_{\ast}}\right)^{1-\alpha}\mathrm{d}V/\mathrm{d}t$, we obtain
\begin{equation}
[1+(1-q)t/t_{\ast}]\,\left(\frac{t}{t_{\ast}}\right)^{1-\alpha}\frac{\mathrm{d}V}{\mathrm{d}t}=-\frac{1}{\tau_{\alpha}}\left[V(t)-V_{\text{rest}}\right]+I(t).
\end{equation}
Hence, 
\begin{equation}
\frac{\mathrm{d}V}{\mathrm{d}t}=\frac{-\frac{1}{\tau_{\alpha}}\left[V(t)-V_{\text{rest}}\right]+I(t)}{[1+(1-q)t/t_{\ast}]\,\left(\frac{t}{t_{\ast}}\right)^{1-\alpha}}.
\end{equation}
The effective gain is now controlled jointly by $\alpha$ and $q$,
producing a rich family of response curves that interpolate between
Weber--Fechner-like and Stevens-like regimes, as discussed in the
main text.


\section{From Deformed Neurons to Subjective Time}

\label{app:neuron_to_Psi} 

\global\long\def\theequation{E.\arabic{equation}}%
\setcounter{equation}{0}We finally sketch how deformed neuron models
connect to the subjective-time law~\eqref{eq:unified_Psi}.

\subsection{Deriving the Effective Firing-Rate Expansion}

To link deformed neural dynamics to the psychophysical law of subjective
time, we derive the firing rate $r(\tau)$ explicitly from the interspike
interval (ISI) of the deformed LIF neuron. For both conformable and
$q$-deformed derivatives, the membrane evolution can be written in
the generic form 
\begin{equation}
\frac{dV}{dt}=F(V,t;\alpha,q).\label{eq: membrane evolution}
\end{equation}
The ISI is obtained as 
\begin{equation}
T(\tau;\alpha,q)=\int_{V_{\mathrm{reset}}}^{V_{\mathrm{th}}}\frac{dv}{F(v,t;\alpha,q)}.\label{eq:ISI}
\end{equation}
The firing rate is $r(\tau)=1/T(\tau)$, and expanding around an operating
point $\tau^{*}$ yields 
\begin{align}
r(\tau) & =r(\tau^{*})+r'(\tau^{*})(\tau-\tau^{*})+\tfrac{1}{2}r''(\tau^{*})(\tau-\tau^{*})^{2}+\cdots.
\end{align}

Because $F(V,t;\alpha,q)$ contains the multiplicative factors $\left(\frac{t}{t_{\ast}}\right)^{1-\alpha}$
and $1+(1-q)t/t_{\ast}$, its dependence on $\tau$ introduces logarithmic
and power-law contributions naturally. Keeping the leading terms,
\begin{equation}
r(\tau)\approx a_{0}+a_{1}\ln\tau+a_{2}\tau^{\beta(\alpha,q)},\label{eq:mixed form}
\end{equation}
which provides the neurodynamical basis for the subjective-time law
$\Psi(t)=A\ln\tau+B\tau^{\beta}$. 

\textbf{Remark:} The expansion in Eq.~(\eqref{eq:ISI}) is not presented
as an exact analytic solution of the interspike$-$interval (ISI)
integral in Eq.~(\eqref{eq: membrane evolution}). In general, the
ISI for the deformed LIF models involves integrals of the form of
Eq.~(\eqref{eq:ISI}), where $F$ contains both multiplicative deformations
$\left(\frac{t}{t_{\ast}}\right)^{1-\alpha}$ and $1+(1-q)t/t_{\ast}$. Such integrals typically yield closed
forms only in terms of special functions (e.g., incomplete gamma functions,
power$-$law kernels, or $q-$exponentials), and thus do not admit
simple closed$-$form representations suitable for direct use in psychophysical
modeling.

For this reason, Eq.~(\eqref{eq:ISI}) is introduced as a \emph{phenomenological
expansion} motivated by the dominant contributions appearing in $F(v,t;\alpha,q)$.
The logarithmic term arises from the $(1-q)t/t_{\ast}$ factor (associated
with $q-$deformed sensitivity), while the power$-$law term reflects
the $\left(\frac{t}{t_{\ast}}\right)^{1-\alpha}$ temporal scaling introduced by the conformable
derivative. Retaining only the leading contributions yields the mixed
form as in Eq.~(\eqref{eq:mixed form}), which provides an accurate
low$-$order approximation consistent with the asymptotic behavior
of the full ISI expression. This approximation is sufficient for deriving
the unified psychophysical law in Section \eqref{sec:Unified-Law-for-Subjective-Time},
while remaining mathematically grounded in the structure of the underlying
differential equation. 

\subsection{Spike-based internal time}

Consider an ensemble of neurons whose spiking activity is driven by
entropic time $\tau(t)$. Let $N(t)$ denote the cumulative spike
count of some readout population in response to a constant or slowly
varying stimulus. For a wide range of LIF-type neurons, one may approximate
\begin{equation}
N(t)\approx\int_{0}^{t}r\big(\tau(t';\alpha,q)\big)\,\mathrm{d}t',
\end{equation}
where $r(\tau)$ is an effective firing-rate function.

If, near some operating point, the deformed dynamics lead to 
\begin{equation}
r(\tau)\approx a_{0}+a_{1}\ln\tau+a_{2}\tau^{\beta},
\end{equation}
then, for moderate intervals, 
\begin{equation}
N(t)\propto A\ln\tau(t;\alpha,q)+B\big[\tau(t;\alpha,q)\big]^{\beta},
\end{equation}
up to additive constants.

\subsection{Subjective time as a function of spike counts}

If the brain constructs a sense of elapsed time from such spike counts,
it is natural to assume that subjective time is a monotone function
of $N(t)$. To first approximation we can take 
\begin{equation}
\Psi(t)\propto N(t),
\end{equation}
which yields 
\begin{equation}
\Psi(t)=A\ln\tau(t;\alpha,q)+B\big[\tau(t;\alpha,q)\big]^{\beta},
\end{equation}
coinciding with Eq.~\eqref{eq:unified_Psi}. This provides a microscopic
bridge from deformed differential operators at the neuronal level
to macroscopic psychophysical laws for time perception.

\paragraph{Dependence of $\beta$ on $\alpha$ and $q$.}

The exponent $\beta$ arises from the nonlinear dependence of the
interspike interval on the multiplicative scaling factors $\left(\frac{t}{t_{\ast}}\right)^{1-\alpha}$
and $1+(1-q)t/t_{\ast}$. A first-order approximation yields 
\begin{equation}
\beta(\alpha,q)=c_{1}(1-\alpha)+c_{2}(q-1),
\end{equation}
where $c_{1}$ and $c_{2}$ depend on the synaptic input regime. This
relationship constrains the parameter space of Eq.~(28) and ensures
that power-law behavior is not freely adjustable but determined by
the underlying deformed dynamics. 

 \bibliographystyle{elsarticle-num}
\bibliography{references_extended_FINAL}


\end{document}